\documentclass[a4paper,11pt]{article}
\pdfoutput=1 

\usepackage{jcappub} 

\usepackage[T1]{fontenc} 
\usepackage{natbib}
\usepackage{amsmath,amssymb,bm}
\usepackage{enumitem,units,multirow}
\usepackage{hyperref}
\usepackage{xcolor}

\newcommand{\ie}{{i.e.}}

\newcommand{\eg}{{e.g.}}

\newcommand{\py}{\small\ttfamily}
\newcommand{\ii}{{\rm i}}
\newcommand{\ee}{{\rm e}}
\newcommand{\cov}{{\rm Cov}}

\title{Beyond Limber: Efficient computation of angular power spectra for galaxy clustering and weak lensing}

\author[1]{Xiao Fang,}
\author[1,2]{Elisabeth Krause,}
\author[1]{Tim Eifler,}
\author[3,4]{and Niall MacCrann}
\affiliation[1]{Department of Astronomy and Steward Observatory, University of Arizona, 933 N Cherry Ave, Tucson, AZ 85719}
\affiliation[2]{Department of Physics, University of Arizona, 1118 E Fourth Str, AZ 85721}
\affiliation[3]{Center for Cosmology and Astro-Particle Physics, The Ohio State University, Columbus, OH 43210, USA}
\affiliation[4]{Department of Physics, The Ohio State University, Columbus, OH 43210, USA}
\emailAdd{xfang@email.arizona.edu}

\abstract{Angular two-point statistics of large-scale structure observables are important cosmological probes. To reach the high accuracy required by the statistical precision of future surveys, some of these statistics may need to be computed without the commonly employed Limber approximation; the exact computation however requires integration over Bessel functions, and a brute-force evaluation is slow to converge. We present a new method based on our generalized FFTLog algorithm for the efficient computation of angular power spectra beyond the Limber approximation. The new method significantly simplifies the calculation and improves the numerical speed and stability. It is easily extended to handle integrals involving derivatives of Bessel functions, making it equally applicable to numerically more challenging cases such as contributions from redshift-space distortions and Doppler effects. We implement our method for galaxy clustering and galaxy-galaxy lensing power spectra. We find that using the Limber approximation for galaxy clustering in future analyses like LSST Year 1 and DES Year 6 may cause significant biases in cosmological parameters, indicating that going beyond the Limber approximation is necessary for these analyses.}


\begin{document}
\maketitle
\flushbottom

\section{Introduction}
\label{sec:intro}

Current and future photometric surveys, such as KiDS\footnote{Kilo-Degree Survey, http://kids.strw.leiden\textbf{}univ.nl/}, HSC\footnote{Hyper Suprime-Cam, https://www.naoj.org/Projects/HSC/}, DES\footnote{Dark Energy Survey, https://www.darkenergysurvey.org}, LSST\footnote{Large Synoptic Survey Telescope, https://www.lsst.org/}, WFIRST\footnote{Wide Field Infrared Survey Telescope, https://wfirst.gsfc.nasa.gov/}, SPHEREx\footnote{Spectro-Photometer for the History of the Universe, Epoch of Reionization, and Ices Explorer, http://spherex.caltech.edu/}, Euclid\footnote{https://www.euclid-ec.org/}, aim to measure the structure of the late-time Universe at high accuracy, enabling precise analyses of various correlations of tracers (galaxy density, weak lensing, galaxy clusters, etc.). Ultimately these endeavors will provide powerful constraints on the nature of dark energy and tests of theories of gravity on cosmological scales. Analyses of photometric surveys primarily use the angular two-point statistics as cosmological probes. The so-called ``$3\times 2$pt'' analysis, combining the galaxy clustering, the galaxy-galaxy lensing (GGL), and the lensing shear-shear correlation, has become a standard set of probes in DES and part of the baseline analysis of LSST-DESC \cite{2018arXiv180901669T}, while joint analyses with other probes \eg, CMB anisotropies and lensing, cluster clustering and lensing, etc., are also playing important roles or currently being investigated.

The modeling of the angular power spectra of two tracers involves double (spherical) Bessel integrals or their derivatives in the form of \footnote{This form only works for flat cosmologies. For curved cosmologies, the functional forms of $f$ and $w_1,w_2$ are modified, and the spherical Bessel functions are replaced with the hyperspherical Bessel functions (see \eg, \cite{2014JCAP...09..032L}).}
\begin{equation}
    \int d\chi_1\,w_1(\chi_1)\int d\chi_2\,w_2(\chi_2) \int_0^\infty dk\,f(k,z(\chi_1),z(\chi_2))j^{(n_1)}_\ell(k\chi_1)j^{(n_2)}_\ell(k\chi_2)~,
    \label{eq:jj_int_example}
\end{equation}
where $n_1,n_2$ are the orders of the derivatives, $\chi_1,\chi_2$ are the comoving distances, $w_1(\chi_1),w_2(\chi_2)$ are the tracers' selection functions (characterizing the efficiency of observing the tracers or their relevant quantities as functions of redshifts), and $f(k,z(\chi_1),z(\chi_2))$ is a smooth function of the wavenumber $k$ and two tracers' redshifts. The full calculation can be numerically challenging due to the rapidly oscillatory nature of the Bessel functions. The Limber approximation  \cite{1953ApJ...117..134L,1954ApJ...119..655L,1973ApJ...185..413P} simplifies the integrals by approximating the spherical Bessel function as a Dirac delta function located at its first peak, \ie, $j_\ell(x)\simeq\sqrt{\pi/(2\ell+1)}\delta(\ell+1/2-x)$ \citep[\eg,][]{2004PhRvD..69h3524A,2018arXiv181205995C}. The derivatives of the spherical Bessel functions can be written in terms of a few spherical Bessel functions of different orders using the recurrence relations, and then approximated with Delta functions. However, this approximation only works well when (1) the (smooth) selection function has a radial width $\Delta\chi$ much larger than the scales of the tracer's fluctuation modes that we are probing (\ie, $\Delta\chi\gg 1/k\sim \bar{\chi}/\ell$, where $\bar{\chi}$ is the mean distance), \textit{and} (2) the two selection functions largely overlap in redshift, \textit{and} (3) $\ell\gg 1$ (see \eg, \cite{2007A&A...473..711S,2008PhRvD..78l3506L} for more discussions). The first requirement is violated, for example, when the photometric redshifts are sufficiently accurate for galaxy clustering measurements and the tomographic bins are chosen so narrow that the selection function is too narrow. The second requirement is violated, for example, when cross correlating different narrow tomographic bins of galaxy density fields. The third requirement is violated, for example, when modeling wide-angle correlations.

The Limber approximation may not be sufficient for modeling angular two-point statistics in analyses of upcoming surveys for several reasons: (1) The aforementioned requirements are violated when the analyses employ narrower tomography bins enabled by the improved photometric redshift accuracy, or when wide-angle correlations are modeled; (2) the improved constraining power can no longer tolerate the errors introduced by the approximation. One may extend the Limber approximation by including high-order correction terms in the series expansion of the angular power spectra in $1/(\ell+1/2)$, as suggested in \cite{2008PhRvD..78l3506L}. However, its accuracy will still depend on the value of $\ell$, and the widths, shapes, and mean redshifts of the selection functions. Meanwhile, including higher-order terms does not alleviate the divergence of the series expansion at small $\ell$, and the convergence radius is case-dependent.

Several approaches have been developed to efficiently evaluate the double Bessel integral in Eq.~(\ref{eq:jj_int_example}), and Fast Fourier Transform (FFT) techniques usually play a central role. In general, the FFT-based methods perform a power-law decomposition of the smooth part of the integrand, each component integral is then calculated analytically, and finally a re-summation of the components yields the result. This approach derives from the FFTLog algorithm, which has long been used to efficiently evaluate single Bessel or spherical Bessel integrals. In Ref.~\cite{2017JCAP...11..054A,2018PhRvD..97b3504G,2018arXiv180709540S}, the ``1D-FFTLog'' method is developed, which decomposes the double Bessel integral into a series of power-law double Bessel integrals (\ie, $\int_0^\infty dk\,k^{\alpha}j_\ell(k r_1)j_\ell(k r_2)$), each of which has an analytic solution in terms of Gamma functions and hypergeometric functions. However, the evaluations of hypergeometric functions can be numerically challenging, and require specialized methods to improve the speed and stability \cite{2018PhRvD..97b3504G}. At low $\ell$'s, the spherical Bessel functions can be easily decomposed into series of products of polynomials and sine/cosine functions, and then integrated with the FFTLog algorithm \cite{2019arXiv191200065S}. In addition, FFTLog based algorithms have also been used to accelerate the computation of the one-loop and two-loop order nonlinear perturbation theories in cosmology \citep[\eg,][]{2016PhRvD..93j3528S,2016JCAP...09..015M,2017JCAP...02..030F,2017arXiv170809247B,2016PhRvD..94j3530S,2018JCAP...04..030S,2018arXiv181202728S}. {\py AngPow} \cite{2017A&A...602A..72C} takes a different approach and optimizes the quadrature integration by using a Clenshaw-Curtis-Chebyshev algorithm and significantly reducing the number of the sampled values, leading to similar or faster computation depending on the properties of the integrands and the sampling points.

In this paper, we present a novel FFTLog based method for solving the non-Limber integrals. Instead of speeding up the double-Bessel integrals, we simplify the full non-Limber angular power spectrum integral by noting the small contribution from unequal-time nonlinear terms, which leads to a significant reduction of the computation and avoiding the double-Bessel integral. We also extend the original FFTLog algorithm to be able to compute integrals containing derivatives of Bessel functions, which can be used to efficiently compute angular power spectra including redshift-space distortions (RSD) and Doppler effects.

This paper is structured as follows. In \S\ref{sec:nonlimber}, we start with the example of galaxy clustering and introduce our approximation for simplifying the integrals. We show that with this approximation the power spectrum reduces to integrals containing a single Bessel function or its derivative. In \S\ref{sec:fftlog-and-beyond}, we review the FFTLog algorithm for single-Bessel integrals and generalize it for integrals containing a derivative of the Bessel function. In \S\ref{sec:applications}, we apply our method to two types of angular power spectra: galaxy clustering and GGL, and compare our FFTLog results with the traditional quadrature integration (hereafter ``brute-force'') results. In \S\ref{sec:significance}, we investigate the significance of including non-Limber galaxy clustering and GGL power spectra in future survey analyses such as DES Year 6 (DES Y6) and LSST Year 1 (LSST Y1). Finally, we discuss other applications and summarize in \S\ref{sec:discussion}. Some useful special function identities are provided in Appendix \ref{app:identities}. We show more FFTLog and brute-force comparison results in Appendix \ref{app:fft_vs_bf}.

\section{Non-Limber Angular Power Spectra}
\label{sec:nonlimber}
In the linear regime, the angular power spectrum between tracers $a$ at redshift $z_1$ and $b$ at redshift $z_2$ can be written as
\begin{equation}
    C_\ell^{ab}=\frac{2}{\pi}\int_0^\infty\frac{dk}{k}k^3P_\Phi(k)\Delta_\ell^a(k,z_1) \Delta_\ell^b(k,z_2)~,
\label{eq:Cl}
\end{equation}
where $P_\Phi(k)$ is the power spectrum of the primordial curvature perturbations, and $\Delta^a$ and $\Delta^b$ are the transfer functions of the tracers (see \eg, Section 2.4.1 of \cite{2018arXiv181205995C} for a list of examples), each of which contains a spherical Bessel integral or its derivative. On small scales (large $\ell$), the Limber approximation can be applied to reduce the integral.

In \S\ref{subsec:ex-gc} we use galaxy clustering as an example to illustrate the problem. In \S\ref{subsec:powerseparation}, we introduce a method to significantly simplify the problem, avoiding the trouble of computing double Bessel integrals.

\subsection{Example: Galaxy Clustering}
\label{subsec:ex-gc}
Galaxy clustering quantifies correlations between galaxy number density fields (we will focus on auto-correlations in this work). From now on we assume a linear galaxy bias model \footnote{The linear galaxy bias model assumes the galaxy power spectrum is proportional to the matter power spectrum. Here we use the nonlinear matter power spectrum rather than the linear power spectrum.}, but comment on the application to non-linear bias models at the end of \S\ref{subsec:powerseparation}. The galaxy number density transfer function $\Delta^{\rm g}$ has 3 components: a galaxy density contribution $\Delta^{\rm D}$ that is proportional to the dark matter density, a linear contribution from redshift space distortions (RSD) $\Delta^{\rm RSD}$ \citep[\eg,][]{1987MNRAS.227....1K,2007MNRAS.378..852P,2019MNRAS.489.3385T}, and a lensing magnification contribution $\Delta^{\rm M}$ \citep[\eg,][]{1995astro.ph.12001V,1998MNRAS.294L..18M,2008PhRvD..77b3512L,2009PhRvL.103e1301S,2014PhRvD..89b3515L,2020MNRAS.491.4869T}. Following \cite{2018arXiv181205995C} and putting back the speed of light constant $c$, they are given by
\begin{align}
&\Delta_\ell^{\rm D} = \int dz\,n_z(z)b(z)T_\delta(k,z) j_\ell(k\chi(z))~,\\
&\Delta_\ell^{\rm RSD}(k) = \int dz \frac{(1+z)n_z(z)}{H(z)}T_\theta(k,z)j_\ell''(k\chi(z))~,\\
&\Delta_\ell^{\rm M}(k) = -\ell(\ell+1)\int\frac{dz}{cH(z)}W^{\rm M}(z)T_{\phi+\psi}(k,z)j_\ell(k\chi(z))~,
\end{align}
where $z$ is redshift, $n_z$ is the normalized redshift distribution of the galaxies, $b$ is the galaxy linear clustering bias, $T_\delta,T_\theta,T_{\phi+\psi}$ are the transfer function of matter perturbations $\delta$, velocity divergences $\theta$, and the Newtonian-gauge scalar metric perturbations, respectively, and are related to each other by
\begin{equation}
    T_\theta(k,z)=-\frac{H(z)f(z)}{1+z}T_\delta(k,z)~,~~T_{\phi+\psi}(k,z)=-\frac{3H_0^2\Omega_{\rm m}(1+z)}{k^2}T_\delta(k,z)~.
\label{eq:T-relation}
\end{equation}
$\chi$ is the comoving distance, $H$ is the Hubble parameter, $\Omega_{\rm m}$ is the matter density fraction at present, and $W^{\rm M}$ is the lensing magnification window function, given by
\begin{equation}
    W^{\rm M}(z) =\int_z^{\infty} dz' n_z(z')\frac{b_{\rm mag}(z')}{2}\frac{\chi(z')-\chi(z)}{\chi(z)\chi(z')}~,
\end{equation}
where $b_{\rm mag}$ is the magnification bias parameter which encapsulates the linear dependence of the galaxy number density on the convergence at a given point on the sky\footnote{defined such that the galaxy overdensity $\delta_{\rm g}$ is changed by $\Delta\delta_{\rm g}=b_{\rm mag}\kappa\delta_{\rm g}$, where $\kappa$ is the lensing convergence.} (due to the change in solid angle and \eg, luminosity cuts \cite{1995astro.ph.12001V,1998MNRAS.294L..18M,2008PhRvD..77b3512L} and size cuts \cite{2009PhRvL.103e1301S,2014PhRvD..89b3515L}).
The angular power spectrum of galaxy number counts is thus
\begin{equation}
    C_\ell^{\rm gg} = \frac{2}{\pi}\int_0^\infty\frac{dk}{k}k^3P_\Phi(k)\Delta_\ell^{\rm g}(k,z_1)\Delta_\ell^{\rm g}(k,z_2)~,
\label{eq:Cl-gg}
\end{equation}
where $\Delta_\ell^{\rm g}=\Delta_\ell^{\rm D}+\Delta_\ell^{\rm RSD}+\Delta_\ell^{\rm M}$.

The expansion of the product of $\Delta^{\rm g}_\ell$'s leads to integrals containing two Bessel functions and their derivatives. For example, the ``DD'' component is
\begin{equation}
    C_\ell^{\rm DD} =\frac{2}{\pi}\int dz_1\,n_z(z_1)b(z_1)\int dz_2\,n_z(z_2)b(z_2)\int_0^\infty\frac{dk}{k}k^3 P_\delta(k,z_1,z_2)j_\ell(k\chi(z_1))j_\ell(k\chi(z_2))~,
\label{eq:Cl-DD}
\end{equation}
where $P_\delta(k,z_1,z_2)= P_\Phi(k)T_\delta(k,z_1)T_\delta(k,z_2)$ is the matter power spectrum across two redshifts $z_1,z_2$, while projected power spectra involving the RSD have integrands containing $j_\ell j_\ell''$ or $j_\ell'' j_\ell''~$. A brute-force way of computing the non-Limber angular power spectra is to calculate the ``double Bessel'' integrals for a grid of $(z_1,z_2)$ and then perform the outer $z_1$ and $z_2$ integrals. Quadrature integration is known to be slow and numerically unstable due to the rapid oscillations of, and the beatings between, the spherical Bessel functions. Various efficient algorithms have been developed for the double Bessel integrals containing $j_\ell j_\ell$, as described in \S\ref{sec:intro}. An efficient algorithm for $j_\ell j_\ell''$ or $j_\ell'' j_\ell''$ has been discussed in Ref.~\cite{2018arXiv180709540S}, where integration by parts is used to move the derivative operator from the Bessel functions to the smooth part of the integrand.

\subsection{Reduction to 1D Bessel Integrals}
\label{subsec:powerseparation}
We now describe a generic and efficient method to calculate the non-Limber integrals without the aforementioned FFTLog methods for double-Bessel integrals.

This method requires the redshift dependence of the matter power spectrum in the non-Limber integral to factorize, hence we adopt a separation of the linear part $P_{\rm lin}(k,z_1,z_2)$ and nonlinear contribution $(P_\delta-P_{\rm lin})(k,z_1,z_2)$ of the power spectrum. The redshift evolution of the linear part is simply a scaling by the growth factor $G(z)$, \ie,
\begin{equation}
    P_{\rm lin}(k,z_1,z_2) = P_{\rm lin}(k,0)G(z_1)G(z_2)~,
\label{eq:Plin}
\end{equation}
where $P_{\rm lin}(k,0)= P_{\rm lin}(k,z=0)$ is the linear matter power spectrum at present (we will comment at the end of this subsection on scale dependent growth).

The nonlinear contribution is significant only on small scales where the Limber approximation is sufficiently accurate, as we demonstrate later in this subsection. This separation avoids the computation of the unequal-time nonlinear power spectrum $P_\delta(k,z_1,z_2)$ (\eg, \cite{2017PhRvD..95f3522K,2019arXiv190502078C}). Applying this separation to the example of $C_\ell^{\rm DD}$ and with the monotonic  function $\chi(z)$, we can rewrite Eq.~(\ref{eq:Cl-DD}) as
\begin{align}
    C_\ell^{\rm DD} &= \frac{2}{2\ell+1}\int_0^\infty dk\,[\tilde{\Delta}^{\rm D}(\chi_\ell)]^2 [P_\delta(k,z(\chi_\ell))-P_{\rm lin}(k,z(\chi_\ell))]\nonumber\\
    &+\frac{2}{\pi}\int d\chi_1\,\tilde{\Delta}^{\rm D}(\chi_1)G(z(\chi_1))\int d\chi_2\,\tilde{\Delta}^{\rm D}(\chi_2)G(z(\chi_2))\int_0^\infty\frac{dk}{k}k^3 P_{\rm lin}(k,0)j_\ell(k\chi_1)j_\ell(k\chi_2),
\label{eq:Cl-DD_rewrite}
\end{align}
where $\tilde{\Delta}^{\rm D}(\chi)=\frac{1}{c}n_z(z(\chi))b(z(\chi))H(z(\chi))$ and $\chi_\ell=(\ell+1/2)/k$. Now we can avoid computing the double-Bessel integral by switching the order of integration in the second term, \ie,
\begin{equation}
    \frac{2}{\pi}\int_0^\infty\frac{dk}{k}k^3 P_{\rm lin}(k,0)\left[\int\frac{d\chi_1}{\chi_1}\chi_1 \tilde{\Delta}^{\rm D}(\chi_1)G(z(\chi_1))j_\ell(k\chi_1)\right]\left[\int\frac{d\chi_2}{\chi_2}\chi_2 \tilde{\Delta}^{\rm D}(\chi_2)G(z(\chi_2))j_\ell(k\chi_2)\right].
\label{eq:faster}
\end{equation}
The two integrals in the square brackets are Hankel transforms that can be efficiently calculated with the original FFTLog algorithm. The outer $k$ integral can be computed with quadrature integration. In total, this method involves four FFTs (each Hankel transform takes 1 FFT to get Fourier coeffients and 1 FFT to sum up, $\mathcal{O}(N\log N)$, see \S\ref{subsec:fftlog}) and one quadrature ($\mathcal{O}(N)$).

For correlations involving RSD, we obtain integrals in the form of Eq.~(\ref{eq:faster}) except that either or both of the spherical Bessel functions are replaced by their second derivatives, where a modified FFTLog algorithm is needed, which will be introduced in \S\ref{sec:fftlog-and-beyond}.

The Limber approximation of the nonlinear correction term $(P_\delta-P_{\rm lin})$ in Eq.~(\ref{eq:Cl-DD_rewrite}) is sufficiently accurate in realistic cases. Assuming the nonlinear contribution is important for Fourier modes $k\geq k_{\rm cr}$, these modes are projected onto the sky and mostly contribute to angular modes $\ell\simeq k\bar{\chi}\geq k_{\rm cr}\bar{\chi}$, where $\bar{\chi}$ is the mean comoving distance of the selection function. In the case of auto-correlation, based on our discussion in \S\ref{sec:intro} and \cite{2008PhRvD..78l3506L}, the Limber approximation is accurate if $\ell\gg \bar{\chi}/\Delta\chi$, where $\Delta\chi$ is the width of the selection function. Thus, as long as $k_{\rm cr}\gg 1/\Delta\chi$, it is safe to apply the Limber approximation to the nonlinear contribution\footnote{For two bins peaked at unequal $\bar{\chi}_1,\bar{\chi}_2$ with comoving radial widths $\Delta \chi_1=\Delta \chi_2$, the criterion (see \cite{2008PhRvD..78l3506L}) becomes $k_{\rm cr}\gg |\bar{\chi}_1-\bar{\chi}_2|/(\Delta\chi_1\Delta\chi_2)$.}. For example, for a narrow tomographic bin centered at $\bar{\chi}=2000\,$Mpc (redshift $\sim 0.5$), with a redshift width $\Delta z=0.2$, \ie, a distance width $\Delta\chi\sim 660\,$Mpc, the criterion becomes $k_{\scriptscriptstyle\rm cr}\gg 1.5\times 10^{-3}\,$Mpc$^{-1}$. The fractional difference between $P_\delta$ and $P_{\rm lin}$ is well within 1\% for $k<0.01 h/$Mpc and within $\sim$5\% for $k<0.2 h$/Mpc, implying that the criterion $k_{\rm cr}\gg 1/\Delta\chi_{\rm max}$ is well satisfied in realistic cases.

We test the approximation quantitatively in \S\ref{subsubsec:test_approx}, and extend our algorithm to the scale-dependent growth in \S\ref{subsubsec:scale-dep}.

\subsubsection{Accuracy of Limber Approximation for Separable Power Spectra}\label{subsubsec:test_approx}
The Limber approximation only picks up the contributions from the equal-time matter power spectra. To assess the impact of the unequal-time nonlinear matter power spectrum $P_\delta(k,z_1,z_2)$, we define a ``scaled'' nonlinear matter power spectrum $P_\delta^{\rm (scaled)}(k,z_1,z_2)$ which follows the linear growth scaling $P_\delta^{\rm (scaled)}(k,z_1,z_2) = P_\delta(k,0)G(z_1)G(z_2)$, where $P_\delta(k,0)= P_\delta(k,z=0)$. Similar to Eq.~(\ref{eq:Cl-DD_rewrite}), we can split $C_\ell^{\rm DD}$ into two parts and use the Limber approximation on the first term, \ie
\begin{align}
    C_\ell^{\rm DD} &= \frac{2}{2\ell+1}\int_0^\infty dk\,[\tilde{\Delta}^{\rm D}(\chi_\ell)]^2 [P_\delta(k,z(\chi_\ell))-P_\delta(k,0)(G(z(\chi_\ell)))^2]\nonumber\\
    &+\frac{2}{\pi}\int d\chi_1\,\tilde{\Delta}^{\rm D}(\chi_1)G(z(\chi_1))\int d\chi_2\,\tilde{\Delta}^{\rm D}(\chi_2)G(z(\chi_2))\int_0^\infty\frac{dk}{k}k^3 P_\delta(k,0)j_\ell(k\chi_1)j_\ell(k\chi_2),
\label{eq:Cl-DD_2}
\end{align}
We denote the $C_\ell^{\rm DD}$ from this alternative splitting as $C_\ell^{\rm scaled}$.

Since $P_\delta^{\rm (scaled)}(k,z_1,z_2)>P_\delta(k,z_1,z_2)>P_{\rm lin}(k,z_1,z_2)$ for any redshift $z_1,z_2>0$,\footnote{This is true for the geometric approximation, $P_\delta(k,z_1,z_2)\simeq\sqrt{P_\delta(k,z_1)P_\delta(k,z_2)}$, and we have verified for the {\py HaloFit} and SPT 1-loop nonlinear power spectrum. For $z_1\neq z_2$, with the more accurate Zel'dovich approximation, $P_\delta(k,z_1,z_2)$ cuts off exponentially at $k\gtrsim k_{\rm NL}/|G(z_1)-G(z_2)|$ \cite{2019arXiv190502078C}, where the nonlinear scale is defined as $k_{\rm NL}^{-2}=(12\pi^2)^{-1}\int_0^\infty P_{\rm lin}(k')dk'$. The cut-off, effective at large $k$ (typically $\gg 0.1h/$Mpc), suppresses the unequal-time contributions, hence improving the accuracy of the Limber approximation.} $C_\ell^{\rm scaled}$ overestimates the unequal-time nonlinear contribution from the matter power spectrum. Thus, the fractional difference $\epsilon_\ell$ between $C_\ell^{\rm DD}$ in Eq.~(\ref{eq:Cl-DD_rewrite}) and $C_\ell^{\rm scaled}$, \ie, $\epsilon_\ell = |C_\ell^{\rm DD}/C_\ell^{\rm scaled} -1|$, provides an upper-bound on the error from using the Limber approximation for $(P_\delta-P_{\rm lin})(k,z_1,z_2)$.

\begin{figure}
    \centering
    \includegraphics[width=0.7\textwidth]{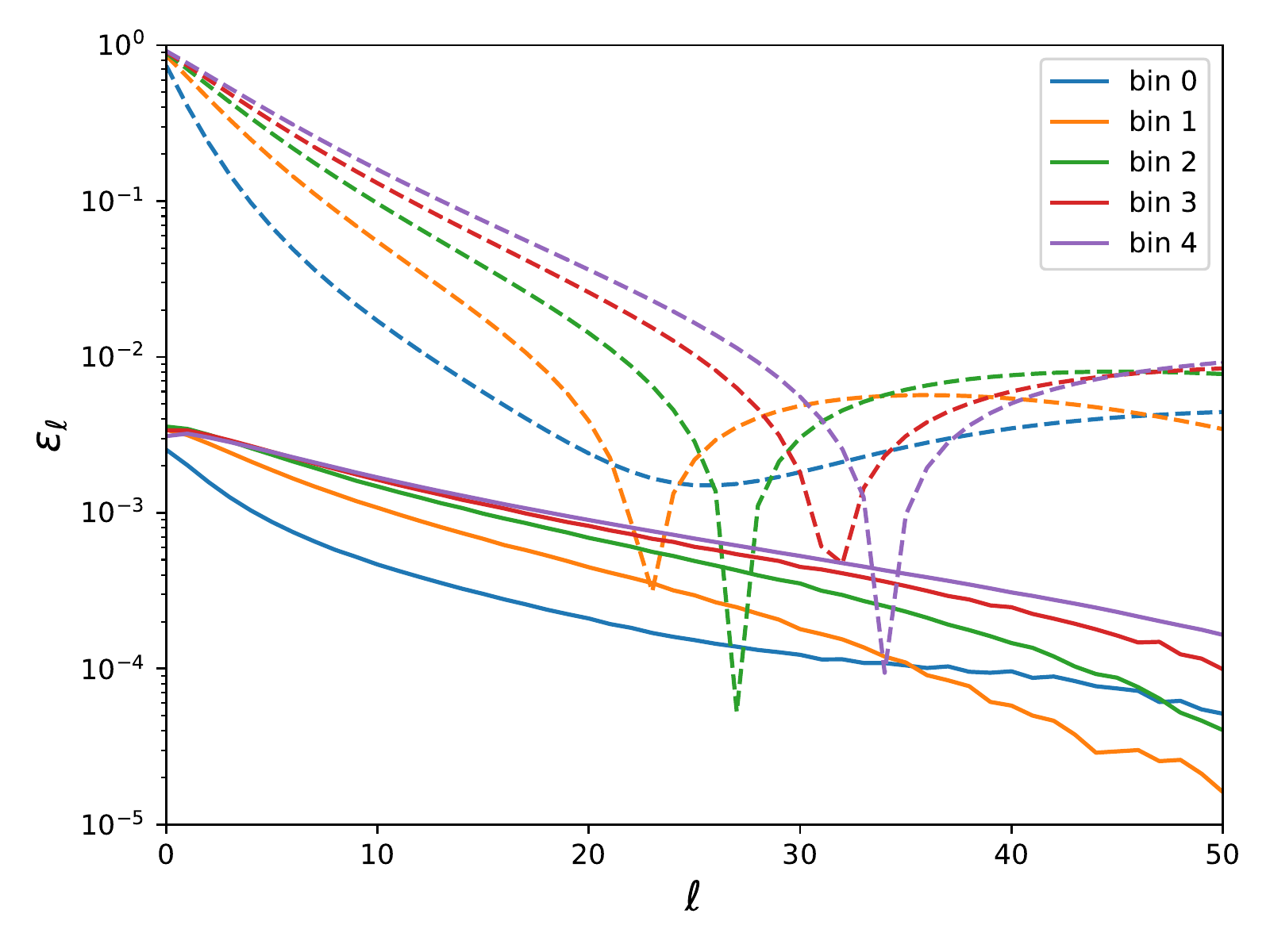}
    \caption{The solid lines show the fractional differences $\epsilon_{\ell}$ between $C_\ell^{\rm DD}$ in Eq.~(\ref{eq:Cl-DD_rewrite}) and $C_\ell^{\rm scaled}$ in Eq.~(\ref{eq:Cl-DD_2}) for the 5 lens tomographic bins of LSST Y1 described in \S\ref{subsubsec:samples}. Bins 0-5 are ordered with increasing redshifts. $C_\ell^{\rm scaled}$ overestimates the unequal-time nonlinear contribution from the matter power spectrum. Thus, $\epsilon_\ell$ is an over-estimation of the impact of the Limber approximation of the first term in Eq.~(\ref{eq:Cl-DD_rewrite}). We also plot in dashed lines the fractional differences between Eq.~(\ref{eq:Cl-DD_rewrite}) and the full Limber version of $C_\ell^{\rm DD}$.}
    \label{fig:test_power_sep}
\end{figure}

In Figure (\ref{fig:test_power_sep}), we show in solid lines the fractional differences $\epsilon_{\ell}$ between $C_\ell^{\rm DD}$ and $C_\ell^{\rm scaled}$ for $\ell\leq 50$ and for the 5 lens tomographic bins of LSST Y1 described in \S\ref{subsubsec:samples}, and find the errors are all below $\sim$0.35\%. We also plot in dashed lines the fractional differences between Eq.~(\ref{eq:Cl-DD_rewrite}) and the full Limber version of $C_\ell^{\rm DD}$, showing that using the Limber approximation of $C_\ell^{\rm DD}$ results in a $\sim 1\%$ error up to $\ell=50$.

The high accuracy of the Limber approximation for the nonlinear contribution also suggests that a similar split can be applied to perturbative galaxy clustering power spectra, where the non-Limber integral is then carried out over the leading order term, and higher-order terms contribute only in the Limber approximation.

\subsubsection{Approximations for Scale-Dependent Growth}\label{subsubsec:scale-dep}
The separability of the redshift-dependence in the linear power spectrum fails when the growth is scale dependent \eg, when there are massive neutrinos. However, for the case where the Limber approximation breaks due to narrow tomographic bins, the scale-dependent contribution to the growth should be small within each bin, and our method can be extended to this case. After performing the linear and nonlinear power spectrum separation as in \S\ref{subsec:powerseparation}, we again apply the Limber approximation for the nonlinear correction part, while the linear part is inseparable. We define scale-dependent growth factor as $G(k,z)=\sqrt{P_{\rm lin}(k,z)/P_{\rm lin}(k,0)}$. Thus, the equal-time linear power spectrum at redshift $z$ within some narrow bin with a mean value $\bar{z}$ is given by
\begin{equation}
    P_{\rm lin}(k,z) = P_{\rm lin}(k,\bar{z})\left[G(k,z)/G(k,\bar{z})\right]^2\simeq P_{\rm lin}(k,\bar{z})[g_{\rm eff}(z,\bar{z})]^2~,
\end{equation}
where in the second step we neglect the scale-dependent growth within the narrow redshift bin and approximate the growth factor ratio as an effective function of $z$, $g_{\rm eff}(z,\bar{z})$, within the bin. Similarly, the unequal-time linear power spectrum of two narrow bins (with means $\bar{z}_1,\bar{z}_2$) is given by
\begin{align}
    P_{\rm lin}(k,z_1,z_2) &=P_{\rm lin}(k)G(k,z_1)G(k,z_2)=P_{\rm lin}(k)G(k,\bar{z}_1)G(k,\bar{z}_2)\frac{G(k,z_1)G(k,z_2)}{G(k,\bar{z}_1)G(k,\bar{z}_2)}\nonumber\\
    &\simeq\sqrt{P_{\rm lin}(k,\bar{z}_1)P_{\rm lin}(k,\bar{z}_2)}g_{\rm eff}(z_1,\bar{z}_1)g_{\rm eff}(z_2,\bar{z}_2)~.
\end{align}
Thus, for given narrow selection functions, the $(k,z_1,z_2)$-dependence can be approximated as separable and the integral can be solved in the same manner as before.

In general, one can always split the selection functions into narrow enough bins to make the approximation sufficiently accurate.

\section{FFTLog and Beyond}\label{sec:fftlog-and-beyond}
The FFTLog algorithm was originally proposed in Ref.~\cite{1978JCoPh..29...35T} to efficiently perform Hankel transforms (\ie, single-Bessel integrals), and was first applied to cosmology in Ref.~\cite{2000MNRAS.312..257H}. In this section, we will first review the procedure of using FFTLog to compute an integral containing one spherical Bessel function (\S\ref{subsec:fftlog}). Then, we generalize the algorithm to integrals with a Bessel function derivative (\S\ref{subsec:beyond-fftlog}), which has applications including RSD.

\subsection{Review of FFTLog}\label{subsec:fftlog}
The FFTLog can be considered as the FFT of a logarithmically sampled integrand. Suppose we want to evaluate the integral
\begin{equation}
    F(r) = \int_0^\infty \frac{dk}{k}f(k)j_\ell(kr)~,
\end{equation}
where $f(k)$ is a smooth input function sampled logarithmically in $k$. The FFTLog method first decomposes $f(k)$ into a series of power-laws, \ie
\begin{equation}
    f(k_q)=\frac{1}{N}\sum_{m=-N/2}^{N/2}c_m k_0^\nu \left(\frac{k_q}{k_0}\right)^{\nu+\ii\eta_m}~,
\label{eq:hankel}
\end{equation}
where $N$ is the sample size of the input function, $\eta_m = 2\pi m/(N\Delta_{\ln k})$, $\nu$ is the bias index, and $\Delta_{\ln k}$ is the linear spacing in $\ln(k)$, \ie, $k_q = k_0\exp(q\Delta_{\ln k})$ with $k_0$ being the smallest value in the $k$ array. The Fourier coefficients satisfy $c_m^* = c_{-m}$ since function $f(k)$ is real, and are computed by discrete Fourier transforming the ``biased'' input function $f(k)/k^\nu$ as
\begin{equation}
    c_m = W_m\sum_{q=0}^{N-1}\frac{f(k_q)}{k_q^\nu}\ee^{-2\pi\ii mq/N}~,
\end{equation}
where $W_m$ is a window function which smooths the edges of the $c_m$ array and takes the form of Eq.~(C.1) in \cite{2016JCAP...09..015M}. This filtering is found to reduce ringing effects.

Each term is now analytically solvable with Eq.~(\ref{eq:app_int_sphj}), \ie
\begin{align}
    F(r)&= \frac{1}{Nr^\nu}\sum_{m=-N/2}^{N/2}c_m k_0^{-\ii\eta_m}r^{-\ii\eta_m} \int_0^\infty\frac{dx}{x}x^{\nu+\ii\eta_m}j_\ell(x)\nonumber\\
    &=\frac{\sqrt{\pi}}{4Nr^\nu}\sum_{m=-N/2}^{N/2}c_m k_0^{-\ii\eta_m}r^{-\ii\eta_m}g_\ell(\nu+\ii\eta_m)~,
\label{eq:Fr}
\end{align}
where the first equality uses change of variable $x=kr$. The function $g_\ell(z)$ is given by
\begin{equation}
    g_\ell(z) = 2^z \frac{\Gamma\left(\frac{\ell+z}{2}\right)}{\Gamma\left(\frac{3+\ell-z}{2}\right)}~,~~-\ell<\Re(z)<2~,
\label{eq:g_l}
\end{equation}
giving the allowed range of bias index $-\ell<\nu<2$.

Finally, assuming that $r$ is logarithmically sampled with the same linear spacing $\Delta_{\ln r}=\Delta_{\ln k}$ in $\ln r$, we can write the last summation in Eq.~(\ref{eq:Fr}) as
\begin{equation}
        F(r_{p}) =  \frac{\sqrt{\pi}}{4 r_{p}^{\nu}}{\rm IFFT}\left[c_{m}^* (k_0r_0)^{\ii\eta_m} g_{\ell}(\nu-\ii\eta_m)\right]~,
\end{equation}
where $r_{p}$ ($p=0,1,\cdots,N-1$) is the $p$-th element in the $r$ array. IFFT stands for the Inverse Fast Fourier Transform. In summary, this method performs two FFT operations, one in computing $c_m$, one in the final summation over $m$. Thus, the total time complexity is $\mathcal{O}(N\log N)$.

\subsection{Beyond FFTLog}
\label{subsec:beyond-fftlog}
One way to solve integrals involving a derivative of the Bessel function, such as
\begin{equation}
    F_n(r)=\int_0^\infty\frac{dk}{k}f(k)j^{(n)}_\ell(kr)~,
\label{eq:Fn-r}
\end{equation}
where superscript $^{(n)}$ stands for the $n$-th derivative, is to rewrite the derivative in terms of several Bessel functions of different orders using recurrence relations. However, it will inevitably increase the number of Hankel transforms if we were to use the original FFTLog method. In this subsection, we generalize the FFTLog method to directly compute this type of integral.

Following the same procedure of power-law decomposition, we can write Eq.~(\ref{eq:Fn-r}) as
\begin{equation}
    F_n(r)=\frac{1}{Nr^\nu}\sum_{m=-N/2}^{N/2}c_m k_0^{-\ii\eta_m}r^{-\ii\eta_m} \int_0^\infty\frac{dk}{k}k^{\nu+\ii\eta_m}j^{(n)}_\ell(k)~.
\end{equation}
Again, the integral for each $m$ has an analytic solution, which can be shown with integration by parts. We write the solution in the same form a with the FFTLog, \ie,
\begin{equation}
    F_n(r)=\frac{\sqrt{\pi}}{4Nr^\nu}\sum_{m=-N/2}^{N/2}c_m k_0^{-\ii\eta_m}r^{-\ii\eta_m}\tilde{g}_\ell(n,\nu+\ii\eta_m)~,
\end{equation}
and its discrete version assuming $\Delta_{\ln r}=\Delta_{\ln k}$,
\begin{equation}
        F_n(r_{p}) =  \frac{\sqrt{\pi}}{4 r_{p}^{\nu}}{\rm IFFT}\left[c_{m}^* (k_0r_0)^{\ii\eta_m} \tilde{g}_{\ell}(n,\nu-\ii\eta_m)\right]~,
\end{equation}
where $\tilde{g}_\ell(n,z) =4\pi^{-1/2}\int_0^\infty dk\,k^{z-1}j^{(n)}_\ell(k)$. For $n=0$, $\tilde{g}_\ell(0,z)=g_\ell(z)$, and for $n=1,2$, it is given by (see Appendix \ref{app:identities})
\begin{align}
    \tilde{g}_\ell(1,z) &= -2^{z-1}(z-1) \frac{\Gamma\left(\frac{\ell+z-1}{2}\right)}{\Gamma\left(\frac{4+\ell-z}{2}\right)}~,~\left(\begin{array}{ll}
    0<\Re(z)<2~,& {\rm for\ }\ell=0 \\
    1-\ell<\Re(z)<2~, & {\rm for\ }\ell\geq 1
    \end{array}\right)~,\\
    \tilde{g}_\ell(2,z) &= 2^{z-2}(z-1)(z-2) \frac{\Gamma\left(\frac{\ell+z-2}{2}\right)}{\Gamma\left(\frac{5+\ell-z}{2}\right)}~,~\left(\begin{array}{ll}
    -\ell<\Re(z)<2~,& {\rm for\ }\ell=0,1 \\
    2-\ell<\Re(z)<2~, & {\rm for\ }\ell\geq 2
    \end{array}\right)~.
\end{align}
We use $\nu=1.01$ for all $\ell$'s (avoiding integer $\nu$ to avoid hitting singularities of the Gamma function). With this generalized FFTLog algorithm, the integral containing one derivative of a spherical Bessel function also takes 2 FFT operations to compute.

\section{Applications}\label{sec:applications}
Equipped with the generalized FFTLog algorithm, we are ready to compute the non-Limber angular power spectra of various tracers. In this section, we first demonstrate this for the galaxy clustering (\S\ref{subsec:gc}). Then, we apply our algorithm to GGL (\S\ref{subsec:GGL}). For each application, we first convert the integrals to a form that is solvable with FFTLog or its extension, then we implement it to compute data vectors based on the LSST Y1 modeling, and compare the results to the results from a ``brute-force'' routine in {\py CosmoLike}\cite{2017MNRAS.470.2100K}.

\subsection{Galaxy Clustering Power Spectrum}\label{subsec:gc}
\paragraph*{Background and Formalism}
In \S\ref{subsec:ex-gc} and \ref{subsec:powerseparation}, we have shown that the ``DD'' component can be written as a nonlinear contribution, solved with the Limber approximation, and a linear part, solvable with the FFTLog. We now consider the full integral of $C_\ell^{\rm gg}$.

With the transfer function relations Eq.~(\ref{eq:T-relation}), the galaxy number count transfer function can be rewritten as
\begin{align}
    \Delta_\ell^{\rm g} &= \int d\chi \bigg[\underbrace{n_z b\frac{H}{c} j_\ell(k\chi)}_{\rm Density}- \underbrace{n_z f \frac{H}{c} j''_\ell(k\chi)}_{\rm RSD} + \underbrace{\frac{3\ell(\ell+1)H_0^2\Omega_{\rm m}(1+z(\chi))}{c^2k^2}W^{\rm M}j_\ell(k\chi)}_{\rm Magnification}\bigg]T_\delta(k,z(\chi))~\nonumber\\
    & = \int d\chi\,S^{\rm g}(k,\chi) T_\delta(k,z(\chi))~,
\end{align}
where $S^{\rm g}(k,\chi)$ is defined as the terms in the square bracket for simplicity, and to make the equation more compact, we do not write out explicitly the $z$ dependences of $n_z(z)$, $b(z)$, $H(z)$, $W^{\rm M}(z)$. Substituting it into Eq.~(\ref{eq:Cl-gg}), we obtain
\begin{align}
    C_\ell^{\rm gg} =&\frac{2}{\pi}\int_0^\infty\frac{dk}{k}k^3\int d\chi_1\,S^{\rm g}(k,\chi_1)\int d\chi_2\,S^{\rm g}(k,\chi_2)P_\delta(k,z(\chi_1),z(\chi_2))~\nonumber\\
    =&\frac{2}{2\ell+1}\int_0^\infty dk(\tilde{\Delta}^{\rm g}(\chi_\ell))^2\left(P_\delta(k,z(\chi_\ell))-P_{\rm lin}(k,z_\ell)\right)\nonumber\\
    &+\frac{2}{\pi}\int_0^\infty\frac{dk}{k}k^3 P_{\rm lin}(k,0)\left[\int\frac{d\chi_1}{\chi_1}\chi_1 S^{\rm g}(k,\chi_1)G(z(\chi_1))\right] \left[\int\frac{d\chi_2}{\chi_2}\chi_2 S^{\rm g}(k,\chi_2)G(z(\chi_2))\right]~.
\end{align}
The second term of the second equality is the linear matter power spectrum term with separable redshift evolution. The integral in each square bracket can be further broken into three or two integrals (collecting terms with the same order of $j_\ell$'s) solvable with FFTLog or its extension in \S\ref{sec:fftlog-and-beyond}. The first term of the second equality is the Limber approximation of the nonlinear contribution, where we reduce $S^{\rm g}(k,\chi)$ to $\sqrt{\pi/(2\ell+1)}\tilde{\Delta}^{\rm g}(\chi)\delta(\ell+1/2-k\chi)$ and then integrate over the $\chi_1,\chi_2$ integrals. For simplicity we define the modified transfer function (with the same dimension as $k$) $\tilde{\Delta}^{\rm g}=\tilde{\Delta}^{\rm D}+\tilde{\Delta}^{\rm RSD}+\tilde{\Delta}^{\rm M}$, whose components are given by (also see Eqs.~(75-77) in \cite{2018arXiv181205995C})
\begin{align}
    \tilde{\Delta}^{\rm D}(\chi_\ell)=&\frac{1}{c}n_z(z(\chi_\ell))b(z(\chi_\ell))H(z(\chi_\ell))~,\\
    \tilde{\Delta}^{\rm RSD}(\chi_\ell)=&\frac{1+8\ell}{(2\ell+1)^2}\frac{f(z(\chi_\ell))}{b(z(\chi_\ell))}\tilde{\Delta}^{\rm D}(\chi_\ell)-\frac{4}{2\ell+3}\sqrt{\frac{2\ell+1}{2\ell+3}}\frac{f(z(\chi_{\ell+1}))}{b(z(\chi_{\ell+1}))}\tilde{\Delta}^{\rm D}(\chi_{\ell+1})~,\\
    \tilde{\Delta}^{\rm M}(\chi_\ell)=&\frac{3\ell(\ell+1)\Omega_{\rm m} H_0^2(1+z(\chi_\ell))}{c^2k^2}W^{\rm M}(\chi_\ell)~,
\end{align}
where the RSD term is derived using the recurrence relations of spherical Bessel functions and approximating the Bessel functions as delta functions.

In our implementation, we perform the $\chi$ integrals ranging from $\chi(z=0.002)$ to $\chi(z=4.0)$ with 1000 logarithmically sampled $\chi$ points, which should cover the entire range of density and RSD contributions and most of the magnification contribution. To further reduce the ringing effect, we zero-pad the integrand by additional $N_{\rm pad}=500$ points on both sides, \ie, making the integrand a size-2000 array. We discard the same number of points $N_{\rm pad}$ on both sides after the transform. We test the convergence of the results by varying the size of the zero-padding.

\begin{figure}
    \centering
    \includegraphics[width=0.7\textwidth]{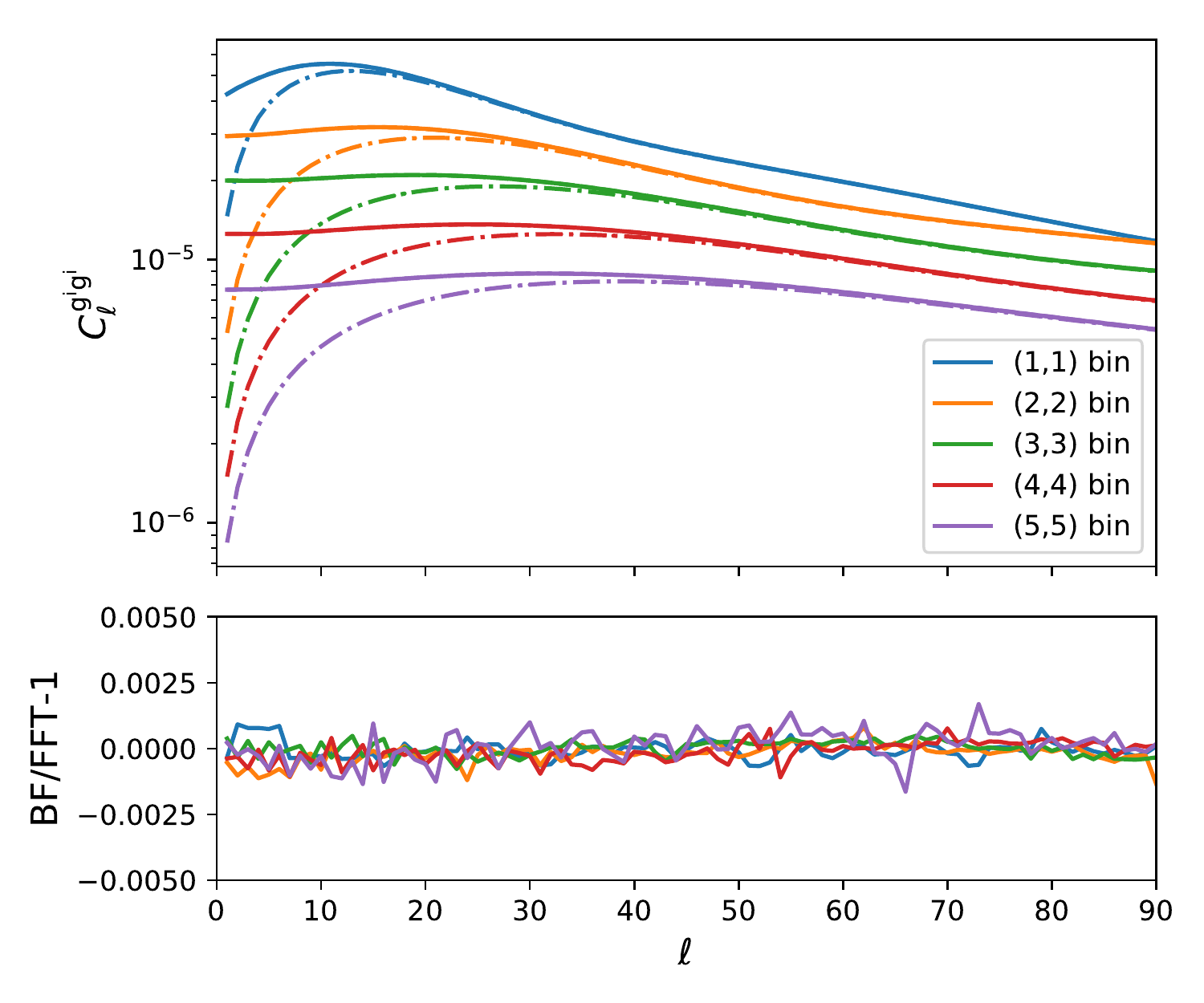}
    \caption{(\textit{Upper panel}:) the non-Limber galaxy clustering auto power spectra $C_\ell^{\rm gg}$'s of the 5 lens tomographic bins in LSST Y1 from the FFT method (solid lines) and the brute-force integration (dashed lines) up to $\ell=90$, along with the Limber results (dash-dotted lines). The FFT and the BF lines nearly overlap with each other. (\textit{Lower panel}:) fractional differences between the BF and the FFT non-Limber results. The fractional differences are all within 0.2\%, consistent with the numerical accuracy of our BF integration.}
    \label{fig:fft_bf_cls}
\end{figure}

\paragraph*{Results}
We compute the galaxy clustering power spectra for LSST Y1, as described in \S\ref{sec:significance}, using our FFTLog method and using a brute-force (BF) quadrature integration. In the upper panel of  Figure \ref{fig:fft_bf_cls} we show in solid lines the non-Limber galaxy clustering power spectra $C_\ell^{\rm gg}$ calculated with our FFT method, and in dashed lines the BF quadrature integration. We also plot the Limber result in dash-dotted lines. In the lower panel, we show the fractional differences between the BF and the FFT non-Limber results (BF/FFT-1). To speed up the BF calculations, we require a 1\% accuracy of the quadrature integration of the non-Limber part and 0.1\% accuracy of the Limber integrals. The fractional differences are all within 0.2\%.

The runtime of performing the $C_\ell^{\rm gg}$ integral for $\ell$ up to 90 is $\sim 0.1\,$s using our FFT method, while it can be about 4$-$9 minutes using our BF implementation, with longer time for higher redshift bins.

\subsection{Galaxy-Galaxy Lensing Power Spectrum}\label{subsec:GGL}
\paragraph*{Background and Formalism}
Galaxy-galaxy lensing (GGL) is the correlation between the shape of background (or \emph{source}) galaxies and foreground (\emph{lens}) galaxy number density. In the weak lensing regime, the observed galaxy shape, $e$ is the sum of an intrinsic (unlensed) component and a shear, $\gamma$, due to gravitational lensing. 
We write their transfer function for the source galaxy shape as $\Delta^{e} = \Delta^{\gamma}+\Delta^{\rm IA}$.

Again with the notations in Ref.~\cite{2018arXiv181205995C}, the lensing effect is characterized by
\begin{equation}
    \Delta_\ell^{\gamma} = -\frac{1}{2}\sqrt{\frac{(\ell+2)!}{(\ell-2)!}}\int\frac{dz}{cH(z)}W^{\rm L}(z)T_{\phi+\psi}(k,z)j_\ell(k\chi(z))~,
\end{equation}
where we have put back the constant $c$ to make $\Delta_\ell^{\rm L}$ dimensionless. The lensing kernel $W^{\rm L}$ is given by
\begin{equation}
    W^{\rm L}(z)=\int_z^\infty dz'\,n_{\rm src}(z')\frac{\chi(z')-\chi(z)}{\chi(z')\chi(z)}~,
\end{equation}
where $n_{\rm src}(z)$ is the redshift distribution of the source galaxies. We use the ``nonlinear linear alignment model'' of IA (\eg, \cite{2001MNRAS.320L...7C,2004PhRvD..70f3526H,2007MNRAS.381.1197H,2011A&A...527A..26J,2015PhR...558....1T,2015JCAP...08..015B,2016MNRAS.456..207K}, but see \cite{2017arXiv170809247B} for limitations), which gives
\begin{equation}
    \Delta_\ell^{\rm IA} = \sqrt{\frac{(\ell+2)!}{(\ell-2)!}}\int dz\,n_{\rm src}(z)A_{\rm IA}(z)T_{\delta}(k,z)\frac{j_\ell(k\chi(z))}{\vert k\chi(z)\vert^2}~,
\end{equation}
where $A_{\rm IA}(z)$ is the (dimensionless) alignment amplitude, defined by\footnote{Our definition of $A_{\rm IA}$ absorbs the typical normalization factors for IA amplitude and redshift evolution, as well as the fraction of aligned galaxies in the sample. It is equivalent to $F(z)$ defined in Eq.~(8) of Ref.~\cite{2017MNRAS.465.1454H}.}
\begin{equation}
    A_{\rm IA} =-\frac{C_1\rho_{\rm cr}\Omega_{\rm m}}{G(z)}A_0\left(\frac{\bar{L}}{L_0}\right)^\beta\left(\frac{1+z}{1+z_0}\right)^{\eta}= -\frac{C_1\rho_{\rm cr}\Omega_{\rm m}}{G(z)}a_{\rm IA}\left(\frac{1+z}{1+z_0}\right)^{\eta}~,
\end{equation}
where we use $C_1\rho_{\rm cr}\simeq 0.0134$, a normalization derived from SuperCOSMOS observations\cite{2004PhRvD..70f3526H,2007NJPh....9..444B}, $z_0,L_0$ are arbitrary pivot values for the power-law scalings of the redshift (with index parameter $\eta$) and luminosity (with index parameter $\beta$) dependences, and $\bar{L}$ is the weighted average luminosity of the source sample. We reduce the number of free parameters by combining free normalization factor $A_0$ and the luminosity dependence into a single free parameter $a_{\rm IA}$. We take $z_0=0.62$ in our analysis, following the DES Year 1 choice in \cite{2018PhRvD..98d3528T}. Combining the two pieces, we obtain the shear transfer function
\begin{align}
    \Delta^e_\ell &=\int d\chi\Bigg\lbrace\sqrt{\frac{(\ell+2)!}{(\ell-2)!}} \bigg[\underbrace{\frac{3H_0^2\Omega_{\rm m}(1+z(\chi))}{2c^2k^2}W^{\rm L}}_{\rm Lensing}+\underbrace{\frac{n_{\rm src} A_{\rm IA}H}{c k^2\chi^2}}_{\rm IA}\bigg]j_\ell(k\chi)\Bigg\rbrace T_\delta(k,z(\chi))~\nonumber\\
    &= \int d\chi\, S^e(k,\chi)T_\delta(k,z(\chi))~,
\end{align}
where $S^e$ is defined as the terms in the curly bracket, and we have again dropped the $z$ dependences of $n_{\rm src}(z),A_{\rm IA}(z),H(z),W^{\rm L}(z)$ for compactness.

Thus, the GGL angular power spectrum $C_\ell^{\rm ge}$ is
\begin{align}
    C_\ell^{\rm ge}=&\frac{2}{\pi}\int_0^\infty\frac{dk}{k}k^3\int d\chi_1\, S^{\rm g}(k,\chi_1)\int d\chi_2\, S^e(k,\chi_2)P_\delta(k,z(\chi_1),z(\chi_2))\nonumber\\
    =&\frac{2}{2\ell+1}\int_0^\infty dk\,\tilde{\Delta}^{\rm g}(\chi_\ell)\tilde{\Delta}^e(\chi_\ell)\left(P_\delta(k,z(\chi_\ell))-P_{\rm lin}(k,z_\ell)\right)\nonumber\\
    &+\frac{2}{\pi}\int_0^\infty\frac{dk}{k}k^3 P_{\rm lin}(k,0)\left[\int\frac{d\chi_1}{\chi_1}\chi_1 S^{\rm g}(k,\chi_1)G(z(\chi_1))\right] \left[\int\frac{d\chi_2}{\chi_2}\chi_2 S^e(k,\chi_2)G(z(\chi_2))\right]~,
\end{align}
where we reduce $S^{e}(k,\chi)$ to $\sqrt{\pi/(2\ell+1)}\tilde{\Delta}^{e}(\chi)\delta(\ell+1/2-k\chi)$ and integrate over $\chi_1,\chi_2$ in the Limber approximation. We also define the modified transfer function (with the same dimension as $k$) $\tilde{\Delta}^e=\tilde{\Delta}^{\gamma}+\tilde{\Delta}^{\rm IA}$ for simplicity, whose components are given by
(also see Eqs.~(78-79) in \cite{2018arXiv181205995C})
\begin{align}
    \tilde{\Delta}^{\gamma}(\chi_\ell)=&\sqrt{\frac{(\ell+2)!}{(\ell-2)!}}\frac{3\Omega_{\rm m} H_0^2(1+z(\chi_\ell))}{2c^2k^2}W^{\rm L}(z(\chi_\ell))~,\\
    \tilde{\Delta}^{\rm IA}(\chi_\ell)=&\sqrt{\frac{(\ell+2)!}{(\ell-2)!}}\frac{n_{\rm src}(z(\chi_\ell))A_{\rm IA}(z(\chi_\ell))H(z(\chi_\ell))}{c(\ell+1/2)^2}~.
\end{align}
Similar to the case of galaxy clustering, the integrals in the linear part can be evaluated with FFTLog and its extension discussed in \S\ref{sec:fftlog-and-beyond}. Our implementation of the $\chi$ integrals follows the same setting as the galaxy clustering, as discussed in \S\ref{subsec:gc}.

\begin{figure}
    \centering
    \includegraphics[width=0.7\textwidth]{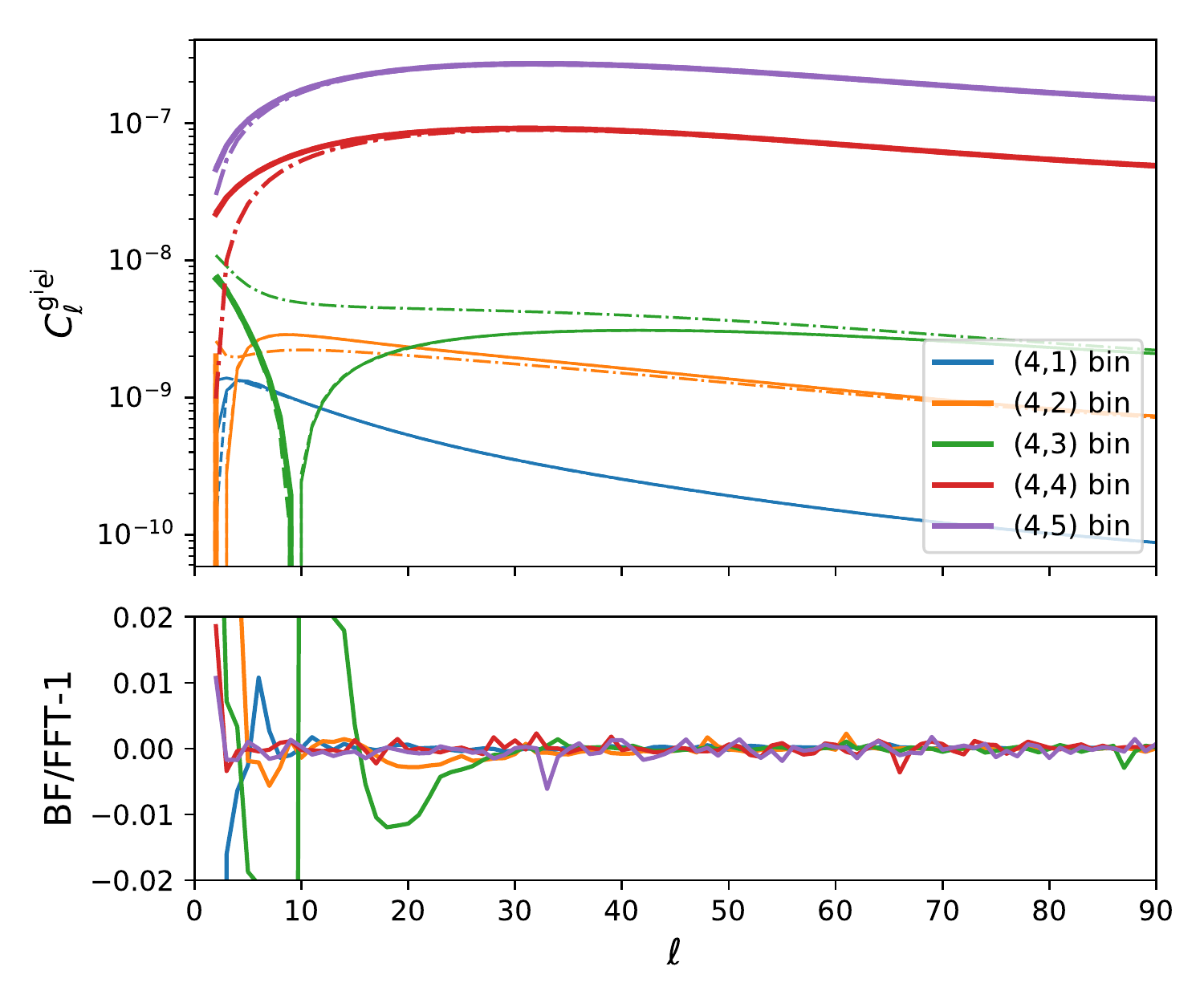}
    \caption{(\textit{Upper panel}:) the non-Limber GGL power spectra $C_\ell^{\rm ge}$ of the 4th lens tomographic bin and all the 5 source bins in LSST Y1 from the FFT method (solid lines) and the brute-force integration (dashed lines) up to $\ell=90$, along with the Limber results (dash-dotted lines). The FFT and the BF lines nearly overlap with each other. (\textit{Lower panel}:) fractional differences between the BF and the FFT non-Limber results. The fractional differences are mostly within 1\%, consistent with the numerical accuracy of our BF integration. Larger errors occur near the zero-crossing in the power spectra.}
    \label{fig:fft_bf_gls}
\end{figure}

\paragraph*{Results}
We compute the GGL power spectra for LSST Y1, as described in \S\ref{sec:significance}, using our FFTLog method and using a brute-force (BF) quadrature integration. In Figure \ref{fig:fft_bf_gls}, on the upper panel we show in solid lines the non-Limber angular power spectra of the 4th lens tomographic bin and all the 5 source bins calculated with our FFT method, in dashed lines with the BF quadrature integration. We also plot the Limber results in dash-dotted lines. On the lower panel, we show the fractional differences between the BF and the FFT non-Limber results (BF/FFT-1). To speed up the BF calculations, we require a 1\% accuracy of the quadrature integration of the non-Limber part and 0.1\% accuracy of the Limber integrals. The fractional differences are mostly within 1\%. Larger errors occur near the zero-crossing in the power spectra. Large differences between the Limber and non-Limber occur in, for example, the cross power spectrum between the 4th lens bin and the 3rd source bin, where two bins largely overlap and the overlapping part of the selection functions are narrow. We test all other 20 lens-source bin combinations of GGL power spectra in Appendix \ref{app:fft_vs_bf}.

The runtime of performing the $C_\ell^{\rm ge}$ integral for $\ell$ up to 90 is $\sim 0.1\,$s using our FFT method, while it varies from 5 minutes to 2.5 hours using our BF implementation, with the longest time for cases where a significant portion of the source galaxy bin is at lower redshifts than the lens galaxy bin.
 
\section{Significance for Future Survey Analyses}\label{sec:significance}
The capability of efficient modeling of angular power spectra beyond the Limber approximation enables us to assess the impact of the Limber approximation on the parameter constraints in future cosmological analyses. We present simulated likelihood analyses of galaxy clustering, GGL and cosmic shear for LSST Y1 and DES Y6.

\subsection{Analysis Ingredients}\label{subsec:ingredients}
\subsubsection{Lens and Source Galaxy Sample Distributions}\label{subsubsec:samples}
\begin{figure}
    \centering
    \includegraphics[width=\textwidth]{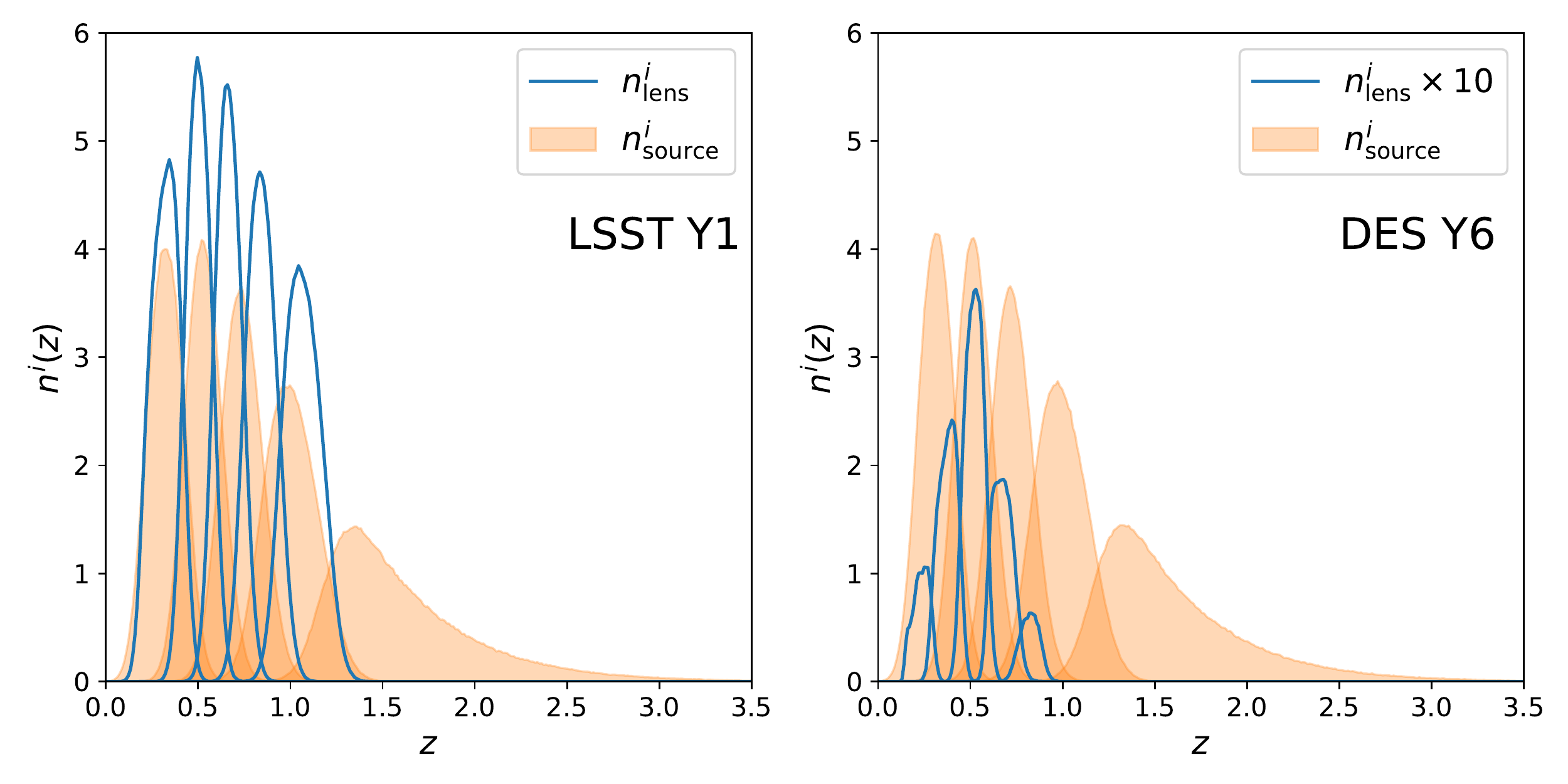}
    \caption{The redshift distributions of the lens and source samples for LSST Y1 (\textit{left}) and DES Y6 (\textit{right}). Each sample is split into 5 bins, as described in \S\ref{subsubsec:samples}.}
    \label{fig:nz}
\end{figure}

We generate the redshift distributions of the lens and source galaxies in LSST Y1 following the DESC Science Requirements Document (DESC SRD, \cite{2018arXiv180901669T}). The LSST Y1 survey will have a survey area of 12.3k deg$^2$ and is expected to measure galaxies with an i-band depth $i_{\rm depth}=25.1\,$mag for the weak lensing (source sample) and galaxies with an i-band limit $i_{\rm lim}=i_{\rm depth}-1=24.1\,$mag for the large-scale structure (lens sample). For the lens sample, we use a parametric redshift distribution (same as Eq.~(5) in Appendix D1.1 of the DESC SRD)
\begin{equation}
	\frac{dN}{dz}\propto z^2\exp[-(z/z_0)^\alpha]~,
\end{equation}
with $(z_0,\alpha)=(0.26,0.94)$, normalized by the effective number density $n_{\rm eff}=18\,$arcmin$^{-2}$. We divide the sample into 5 tomographic bins with equal number of galaxies and convolve each bin with a Gaussian photo-$z$ scatter with $\sigma_z = 0.03(1+z)$. For each tomographic bin $i$, we set their fiducial linear galaxy bias parameters as $b_i = 1.05/G(\bar{z}^i)$, where $\bar{z}^i$ is the mean redshift of the $i$-th bin, and $G(z)$ is the linear growth factor. For the source sample, we use the same parametric form but with $(z_0,\alpha)=(0.191,0.870)$, normalized to $n_{\rm eff}=11.2\,$arcmin$^{-2}$.\footnote{These values for the source sample are the updated version from private communication with Rachel Mandelbaum.} We also divide the source sample into 5 equally populated tomographic bins and convolve each bin with a Gaussian photo-$z$ uncertainty with $\sigma_z=0.05(1+z)$. The distributions of the lens and source tomographic bins are shown on the left panel of Figure \ref{fig:nz}. We assume the galaxy shape noise to be $\sigma_e= 0.26$ per component.

The full DES survey (Y6) has a survey area of 5000 deg$^2$ and is expected to measure galaxies with very similar depth to the LSST Y1. Therefore, we define the source sample to reach $i$-depth of $25.0\,$mag and follow the same parametric redshift distribution. Using fitting formulae from the DESC SRD, we obtain $(z_0,\alpha)=(0.193,0.876)$ and $n_{\rm eff}=10.47\,$arcmin$^{-2}$. Similar to LSST Y1, we divide the source sample into 5 equally populated tomographic bins with $\sigma_z=0.05(1+z)$, but set the shape noise consistent with DES Y1, \ie, $\sigma_e= 0.279$ per component. For the lens sample, we assume a \textsc{redMaGiC} \cite{2016MNRAS.461.1431R} selected sample and split it into 5 bins similar to the DES Y1 analysis \cite{2018PhRvD..98d2006E,2018MNRAS.481.2427C}. We show the distributions of the lens and source tomographic bins on the right panel of Figure \ref{fig:nz}.

\subsubsection{Angular Two-Point Functions}\label{subsubsec:2ptfuncs}
The modeling of the galaxy clustering power spectra and the GGL power spectra has already been shown in \S\ref{subsec:gc} and \ref{subsec:GGL}. We model cosmic shear using the Limber approximation as the lensing efficiency function is very broad\footnote{The impact of the Limber approximation of the cosmic shear power spectrum has been shown sufficiently small in \eg, \cite{2017MNRAS.472.2126K,2017MNRAS.469.2737K,2017JCAP...05..014L}}. Including IA and following the notation in \S\ref{subsec:GGL}, the tomographic cosmic shear power spectra between the source bin $i$ and $j$ can be written as
\begin{equation}
    C_\ell^{e^ie^j} = \frac{2}{2\ell+1}\int_0^\infty dk\,\tilde{\Delta}^{e^i}(\chi_\ell)\tilde{\Delta}^{e^j}(\chi_\ell)P_\delta(k,z(\chi_\ell))~.
\end{equation}
We compute the linear matter power spectrum using the transfer function from Eisenstein and Hu \cite{1998ApJ...496..605E}, and the nonlinear matter power spectrum with {\py HaloFit} \cite{2003MNRAS.341.1311S,2012ApJ...761..152T}. To perform likelihood analyses real space, we calculate the angular two-point correlation functions for  galaxy clustering $w^i(\theta)$, GGL $\gamma_t^{ij}(\theta)$, and cosmic shear $\xi_{+/-}^{ij}(\theta)$, using their relation to angular power spectra on the curved sky:
\begin{align}
    w^i(\theta) &= \sum_\ell \frac{2\ell+1}{4\pi}P_\ell(\cos\theta) C^{{\rm g}^i{\rm g}^i}(\ell)~,\\
    \gamma_t^{ij}(\theta) &= \sum_\ell \frac{2\ell+1}{4\pi\ell(\ell+1)}P^2_\ell(\cos\theta) C^{{\rm g}^i e^j}(\ell)~,\\
    \xi_{\pm}^{ij}(\theta) &= \sum_\ell\frac{2\ell+1}{2\pi\ell^2(\ell+1)^2}[G_{\ell,2}^+(\cos\theta)\pm G_{\ell,2}^-(\cos\theta)] C^{e^i e^j}(\ell)~,
\end{align}
where $\theta$ is the angular separation, $P_\ell$ and $P_\ell^2$ are the Legendre polynomial and the associated Legendre polynomial, $G_{\ell,m}^{+/-}$ are given by Eq.~(4.19) of \cite{1996astro.ph..9149S}. We have used the indices $i,j$ to denote the tomographic bins involved, \ie, $C^{{\rm g}^i{\rm g}^i}(\ell)$ is the galaxy clustering power spectrum of the $i$-th lens tomographic bin, $C^{{\rm g}^i e^j}(\ell)$ is the GGL power spectrum of the $i$-th lens tomographic bin and the $j$-th source bin, and $C^{e^i e^j}(\ell)$ is the tomographic cosmic shear power spectrum of the $i$-th and $j$-th source bins.

We compute all correlation functions in 26 logarithmically spaced angular bins over the range $2.5'<\theta<900'$. For each angular bin $[\theta_{\rm min},\theta_{\rm max}]$, the correlation functions are bin-averaged, \ie, replacing $P_\ell(\cos\theta)$, $P_\ell^2(\cos\theta)$ and $[G_{\ell,2}^+(\cos\theta)\pm G_{\ell,2}^-(\cos\theta)]$ with their bin-averaged functions $\overline{P_\ell}$, $\overline{P^2_\ell}$ and $\overline{G_{\ell,2}^+\pm G_{\ell,2}^-}$ \cite{Friedrich_inprep}, defined by
\begin{align}
    &\overline{P_\ell} = \frac{\int_{\cos\theta_{\rm min}}^{\cos\theta_{\rm max}}dx\,P_\ell(x)}{\cos\theta_{\rm max}-\cos\theta_{\rm min}}=\frac{[P_{\ell+1}(x)-P_{\ell-1}(x)]_{\cos\theta_{\rm min}}^{\cos\theta_{\rm max}}}{(2\ell+1)(\cos\theta_{\rm max}-\cos\theta_{\rm min})}~,\\
    &\overline{P^2_\ell} = \frac{\int_{\cos\theta_{\rm min}}^{\cos\theta_{\rm max}}dx\,P^2_\ell(x)}{\cos\theta_{\rm max}-\cos\theta_{\rm min}} = \frac{[(\ell+\frac{2}{2\ell+1})P_{\ell-1}(x)+(2-\ell)xP_\ell(x) - \frac{2}{2\ell+1}P_{\ell+1}(x)]_{\cos\theta_{\rm min}}^{\cos\theta_{\rm max}}}{\cos\theta_{\rm max}-\cos\theta_{\rm min}}\\
    &\overline{G_{\ell,2}^+\pm G_{\ell,2}^-} = \frac{1}{\cos\theta_{\rm max}-\cos\theta_{\rm min}} \left\lbrace\scriptstyle-\frac{\ell(\ell-1)}{2}\left(\ell+\frac{2}{2\ell+1}\right)P_{\ell-1}(x)-\frac{\ell(\ell-1)(2-\ell)}{2}xP_{\ell}(x)+\frac{\ell(\ell-1)}{2\ell+1}P_{\ell+1}(x)\right.\nonumber\\
    &\ \ \ \ \ \ \ \ \ \ \ \ \ \ \ \ \ \left.\scriptstyle+(4-\ell)\frac{dP_\ell(x)}{dx}+(\ell+2)[x\frac{dP_{\ell-1}(x)}{dx}-P_{\ell-1}(x)]\pm 2(\ell-1)[x\frac{dP_{\ell}(x)}{dx}-P_{\ell}(x)]\mp 2(\ell+2)\frac{dP_{\ell-1}(x)}{dx}\right\rbrace_{\cos\theta_{\rm min}}^{\cos\theta_{\rm max}}~.
\end{align}
Our analysis includes all auto-correlations of the lens bins for the galaxy clustering, all combinations of lens and source bins for the GGL, and all auto- and cross-correlations of the source bins for the cosmic shear. Thus, for each survey, the data vector contains 5 sets of $w(\theta)$, 25 sets of $\gamma_t(\theta)$, 15 sets of $\xi_+(\theta)$ and 15 sets of $\xi_-(\theta)$, each of which has 26 angular bins.

\subsubsection{Systematics}\label{subsubsec:systematics}
Systematic uncertainties are parameterized through nuisance parameters in a similar way to the DES Y1 analysis \cite{2017arXiv170609359K}.
\paragraph{Photometric redshift uncertainties} The uncertainty in the redshift distribution of the $i$-th tomographic bin $n^i(z)$ is modeled by one shift parameter $\Delta_z$ for each bin of the lens and the source samples, \ie, $n^i(z) = \hat{n}^i(z-\Delta_z^i)$, where the index $i$ traverses over all the lens and source bins, and $\hat{n}$ is the estimated redshift distribution as described in \S\ref{subsubsec:samples}. There are 10 shift parameters in total. We take 0 as their fiducial values to generate the simulated data vector, and marginalize over them in the likelihood analyses. For the lens sample in both surveys, we choose a Gaussian prior with $\mu=0,\sigma=0.005(1+\bar{z}^i)$ for each $\Delta_{z,\rm lens}^i$. For the source samples in both surveys, we choose a Gaussian prior with $\mu=0,\sigma=0.002(1+\bar{z}^i)$ for each $\Delta_{z,\rm source}^i$, consistent with the requirements given in \S5.1 and 5.2 of the DESC SRD.

\paragraph{Galaxy bias} We assume a linear bias model and use one parameter for each lens bin. There are 5 parameters in total, whose fiducial values are described in \S\ref{subsubsec:samples} for generating the simulated data vector. In the likelihood analysis, they will be marginalized over with conservative flat priors (between 0.8 and 3).

\paragraph{Lensing magnification bias} Lensing magnification was not included in the baseline model of DES Y1 analysis. We parameterize the effect through one parameter for each lens bin $b_{\rm mag}^i$. For magnitude limited samples, magnification due to lensing by line-of-sight structure can affect the number density of galaxies with observed magnitudes exceeding the magnitude cut (\eg, \cite{1995astro.ph.12001V,1998MNRAS.294L..18M,2008PhRvD..77b3512L}). For LSST Y1, we assume the lens samples are magnitude limited. Using the fitting formulae in Appendix C \footnote{based on the relation between the galaxy redshift distributions and the magnitude limits modeled by \cite{2005MNRAS.363.1329B} using COMBO-17 luminosity functions
for the SDSS r filter \cite{2003A&A...401...73W}.} of \cite{2010A&A...523A...1J} and an $r$-band limit of $24.81$ mag (see Appendix C1 of the DESC SRD and define $r$-band limit for lens sample $r_{\rm lim}=r_{\rm depth}-1=25.81-1$ mag), we obtain $b_{\rm mag}^i = (-0.898, -0.659, -0.403, -0.0704, 0.416)$ for the 5 lens bins. For DES Y6 lens sample, we use luminosity cuts $L/L_*>(0.5,0.5,0.5,1.0,1.0)$ for bins from low to high redshifts, respectively. $L_*$ is the characteristic galaxy luminosity where the power-law form in the Schechter luminosity function cuts off \cite{1976ApJ...203..297S}. To estimate $b_{\rm mag}$, we assume a Schechter luminosity function $N(L)\propto (L/L_*)^\alpha \exp(-L/L_*)~$, where the parameter $\alpha$ characterizes the faint-end slope, $N(L)$ is the number of galaxies per unit luminosity bin. $b_{\rm mag}$ is related to the luminosity function by
\begin{equation}
    b_{\rm mag} = -2-2\left.\frac{d\log_{10}N_{>L}}{d\log_{10}L}\right\vert_{L=L_{\rm lim}} = -2+2L_{\rm lim}\frac{N(L_{\rm lim})}{N_{>L_{\rm lim}}}=-2+2\frac{\left(\frac{L_{\rm lim}}{L_*}\right)^{\alpha+1} e^{-L_{\rm lim}/L_*}}{\Gamma(\alpha+1,L_{\rm lim}/L_*)}~,
\end{equation}
where $L_{\rm lim}$ is the luminosity cut, and $N_{>L}$ is the number of galaxies brighter than luminosity $L$, calculated by integrating the luminosity function. $\Gamma(a,z)$ is the incomplete Gamma function\footnote{The incomplete Gamma function is defined as $\Gamma(a,z)=\int_z^\infty t^{a-1}e^{-t}dt$. See \eg, Eq.~(8.2.2) of DLMF \cite{NIST:DLMF}.}. We assume a constant $\alpha=-0.8$ for all lens bins, leading to $b_{\rm mag}^i = (-0.102, -0.102, -0.102, 1.06, 1.06)$. Note that these values may not represent the exact values for the \textsc{redMaGiC} sample, but merely our approximate choice for generating the fiducial simulated data vector. Although we haven't considered other selection biases, such as the size magnification bias due to the magnification of the angular sizes of source galaxies, the uncertainties of their effects are encapsulated in the 5 nuisance parameters $b_{\rm mag}^i$ and are marginalized over conservative flat priors (between $-3$ and 3) in the likelihood analysis.

\paragraph{Multiplicative shear calibration} We use one shear calibration uncertainty parameter $m^i$ per source bin (5 in total), acting on the cosmic shear and GGL correlation functions such that
\begin{align}
    \xi_{+/-}^{ij}(\theta) &\rightarrow (1+m^i)(1+m^j)\xi_{+/-}^{ij}(\theta)~,\nonumber\\
    \gamma_t^{ij}(\theta) &\rightarrow (1+m^j)\gamma_t^{ij}(\theta)~.
\end{align}
They are marginalized over independently with Gaussian priors ($\mu=0,\sigma=0.005$).

\paragraph{IA} We use the nonlinear linear alignment (NLA) model and parameterize it with two parameters $a_{\rm IA}$ and $\eta$. See \S\ref{subsec:GGL} for detail. Their fiducial values are $a_{\rm IA}=0.5$ and $\eta=0$, and they are both marginalized over independently with conservative flat priors between $-5$ and 5.

\subsubsection{Covariances}\label{subsubsec:covs}
The Fourier space 3x2pt covariances, including the Gaussian part \cite{2004PhRvD..70d3009H}, the connected non-Gaussian part (calculate with the halo model) and the super-sample covariance \cite{2013PhRvD..87l3504T}, are described in Appendix A of \cite{2017MNRAS.470.2100K}.
We calculate the covariances of bin-averaged correlation functions on the curved sky, \ie, for two angular two-point functions, $\Xi,\Theta\in\lbrace w,\gamma_t,\xi_+,\xi_-\rbrace$
\begin{equation}
    \cov(\Xi^{ij}(\theta),\Theta^{km}(\theta')) = \sum_\ell\overline{P^\Xi_\ell}\sum_{\ell'}\overline{P^\Theta_{\ell'}}\,\cov(C_\Xi^{ij}(\ell),C_\Theta^{km}(\ell'))~,
\label{eq:cov_fullsky}
\end{equation}
where $C_{\xi_+}= C_{\xi_-}= C^{ee}$, $C_{\gamma_t}= C^{\rm ge}$, and $C_w= C^{\rm gg}$ in our previous notation, and $i,j,k,m$ are the tomographic bin indices. The bin-averaged weight functions are defined as \cite{Friedrich_inprep}
\begin{align}
    \overline{P^w_\ell}&= \frac{2\ell+1}{4\pi}\overline{P_\ell} = \frac{[P_{\ell+1}(x)-P_{\ell-1}(x)]_{\cos\theta_{\rm min}}^{\cos\theta_{\rm max}}}{4\pi(\cos\theta_{\rm max}-\cos\theta_{\rm min})}~,\\
    \overline{P^{\gamma_t}_\ell}&= \frac{2\ell+1}{4\pi\ell(\ell+1)}\overline{P^2_\ell} = \frac{2\ell+1}{4\pi\ell(\ell+1)} \frac{[(\ell+\frac{2}{2\ell+1})P_{\ell-1}(x)+(2-\ell)xP_\ell(x) - \frac{2}{2\ell+1}P_{\ell+1}(x)]_{\cos\theta_{\rm min}}^{\cos\theta_{\rm max}}}{\cos\theta_{\rm max}-\cos\theta_{\rm min}},\\
    \overline{P^{\xi_{\pm}}_\ell}&= \frac{2\ell+1}{2\pi\ell^2(\ell+1)^2}\overline{G_{\ell,2}^+\pm G_{\ell,2}^-}~,
\end{align}
In practice, we evaluate Eq.~(\ref{eq:cov_fullsky}) up to $\ell_{\rm max}=50000$.

\subsubsection{Angular Scale Cuts}\label{subsubsec:scalecuts}
Limited by our ability to accurately model the non-linearities of the density and galaxy fields on small scales, survey analyses define a set of angular scale cuts, preventing nonlinear modeling limitations from biasing the cosmology results. In accordance with the DESC SRD, we choose a Fourier scale cut $k_{\rm max}=0.3h/$Mpc, which roughly corresponds to a comoving scale $R_{\rm min}=2\pi/k_{\rm max}=21\,$Mpc$/h$.

For the galaxy clustering, we define the angular scale cut 
$\theta_{\rm min}^{w^i}$ for lens tomographic bin $i$ as
\begin{equation}
    \theta_{\rm min}^{w^i} = \frac{R_{\rm min}}{\chi(\bar{z}^i)}~,
\end{equation}
where $\bar{z}^i$ is the mean redshift of galaxies in redshift bin $i$. For our LSST Y1 lens sample, $\theta_{\rm min}^{w^i}$ are $80.88'$, $54.19'$, $42.85'$, $35.43'$, $29.73'$, respectively. For DES Y6, they are $108.3'$, $69.77'$, $52.53'$, $42.37'$, $36.15'$. For the GGL $\gamma_t^{ij}$, we use the same angular cuts as $\theta_{\rm min}^{w^i}$.

For the cosmic shear, we use the Fourier scale cut defined in the DESC SRD, \ie, $\ell<\ell_{\rm max}=3000$, and translate it into the angular cuts for $\xi_{+/-}$ with the first zeros of their corresponding Bessel functions $J_{0/4}$ (in the flat-sky-limit transform), \ie, $\theta_{\rm min}^{\xi_+}=2.4048/\ell_{\rm max}=2.756'$, and $\theta_{\rm min}^{\xi_-}=7.5883/\ell_{\rm max}=8.696'$.

\subsection{Simulated Likelihood Analysis}\label{subsec:like}
The parameter estimation is done by sampling and maximizing the likelihood of the data $\bm{D}$ given a point in cosmological and nuisance parameter space $\bm{p}$,
\begin{equation}
    L(\bm{D}\vert\bm{p})\propto\exp\left(-\frac{1}{2}\left[(\bm{D}-\bm{M}(\bm{p}))^{\rm T}\bm{C}^{-1}(\bm{D}-\bm{M}(\bm{p}))\right]\right)~,
\end{equation}
where $\bm{M}$ is the model vector and $\bm{C}$ is the covariance matrix. We generate the simulated 3x2pt data vector $\bm{D}$ by computing the model vector at the fiducial parameter values and in our fiducial cosmology (standard $\Lambda$CDM with massless neutrinos), with non-Limber modeling of $w$ and $\gamma_t$, and Limber modeling of $\xi_{+/-}$, as described in \S\ref{subsubsec:2ptfuncs}. Throughout our analyses, we use the {\py emcee} sampler \cite{2013PASP..125..306F}.

\subsubsection{The Impact of Limber Approximation}\label{subsubsec:limberimpact}
To evaluate the impact of the Limber approximation in the configurations of LSST Y1 and DES Y6, we fit the data vector with a set of models:
\begin{itemize}
    \item Model (I): A 2x2pt (galaxy clustering + GGL) analysis which uses the Limber approximation to calculate $w,\gamma_t$;
    \item Model (II): A cosmic shear analysis which uses the Limber approximation to calculate $\xi_{+/-}$;
    \item Model (III): A 3x2pt analysis which uses the Limber approximation to calculate all 3x2pt correlations;
    \item Model (IV): A 2x2pt analysis which uses the Limber approximation to calculate $\gamma_t$ and uses our non-Limber method to compute $w$.
\end{itemize}

We summarize the fiducial values and priors of the parameters in Table \ref{tab:params}.

\begin{figure}
    \centering
    \includegraphics[width=0.49\textwidth]{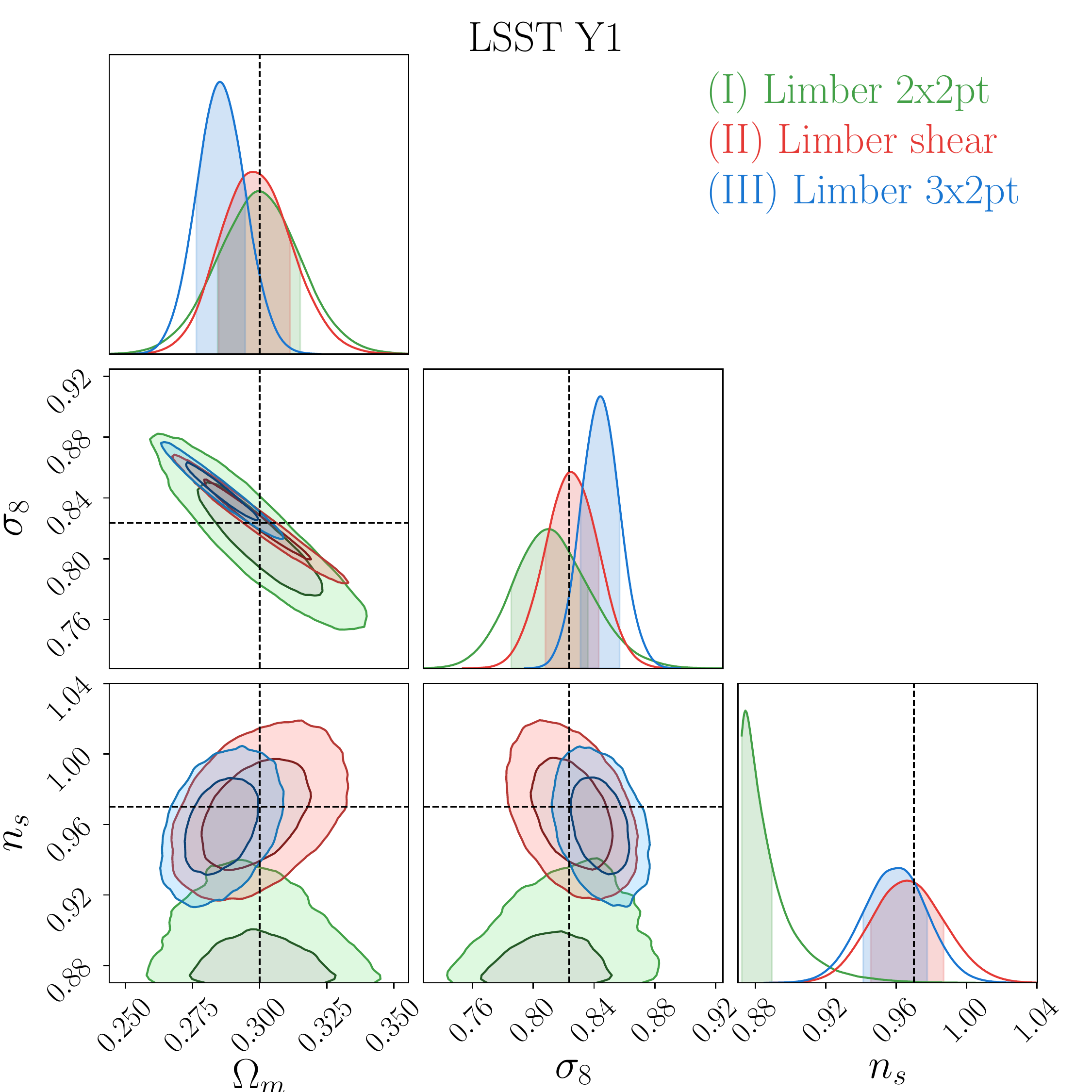}
    \includegraphics[width=0.49\textwidth]{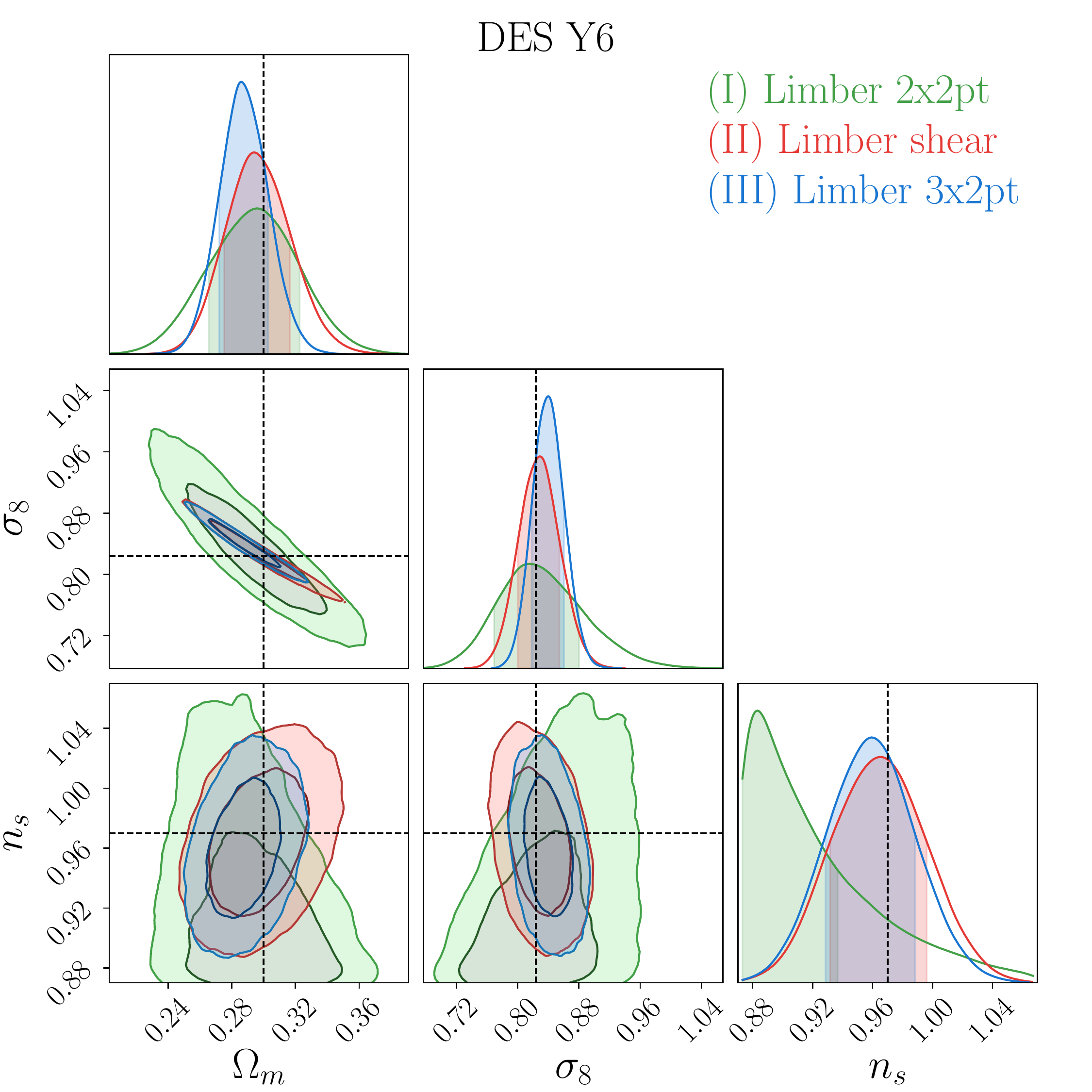}
    \caption{The $1\sigma$ and $2\sigma$ contours of fitting the simulated (non-Limber $w+\gamma_t$, Limber $\xi_\pm$) 3x2pt data vector $\bm{D}$ with Models (I) Limber 2x2pt ($w+\gamma_t$, in green), (II) Limber shear ($\xi_\pm$, in red), (III) Limber 3x2pt (in blue), for LSST Y1 (\textit{left}) and DES Y6 (\textit{right}). The dashed lines mark the fiducial values. In both surveys, Model (I) introduces significant biases to $n_{\rm s}$, while $\Omega_{\rm m},\sigma_8$ are less affected. Model (II) correctly recovers fiducial values of all the parameters as expected. Model (III) gains an improved $n_{\rm s}$ constraint from the cosmic shear, forcing the other two parameters to shift away from their fiducial values by $\sim 1\sigma$.}
    \label{fig:lsst_des_contour_lim_2x2_shear_3x2}
\end{figure}

In Figure \ref{fig:lsst_des_contour_lim_2x2_shear_3x2}, we show the $1\sigma$ and $2\sigma$ contours of the 3 cosmological parameters ($\Omega_{\rm m},\sigma_8,n_{\rm s}$) as well as their marginalized 1D distributions from Models (I-III), in the cases of LSST Y1 and DES Y6, respectively. The Limber approximation modifies the angular power spectra at low-$\ell$, which results in the bias of the spectral index $n_{\rm s}$ for Model (I), the Limber 2x2pt model. For both surveys, $n_{\rm s}$ absorbs most of the cosmological parameter biases. However, since cosmic shear has stronger constraining power on $n_{\rm s}$, the combined 3x2pt result is shifted towards the fiducial $n_{\rm s}$ values, while other parameters may be shifted further away from the fiducial values. For the $\Omega_{\rm m}-\sigma_8$ contour, we see a shift greater than $1\sigma$ for LSST Y1, and a nearly $1\sigma$ shift for DES Y6.

\begin{figure}[!tbp]
  \centering
  \begin{minipage}[t]{0.49\textwidth}
    \includegraphics[width=\textwidth]{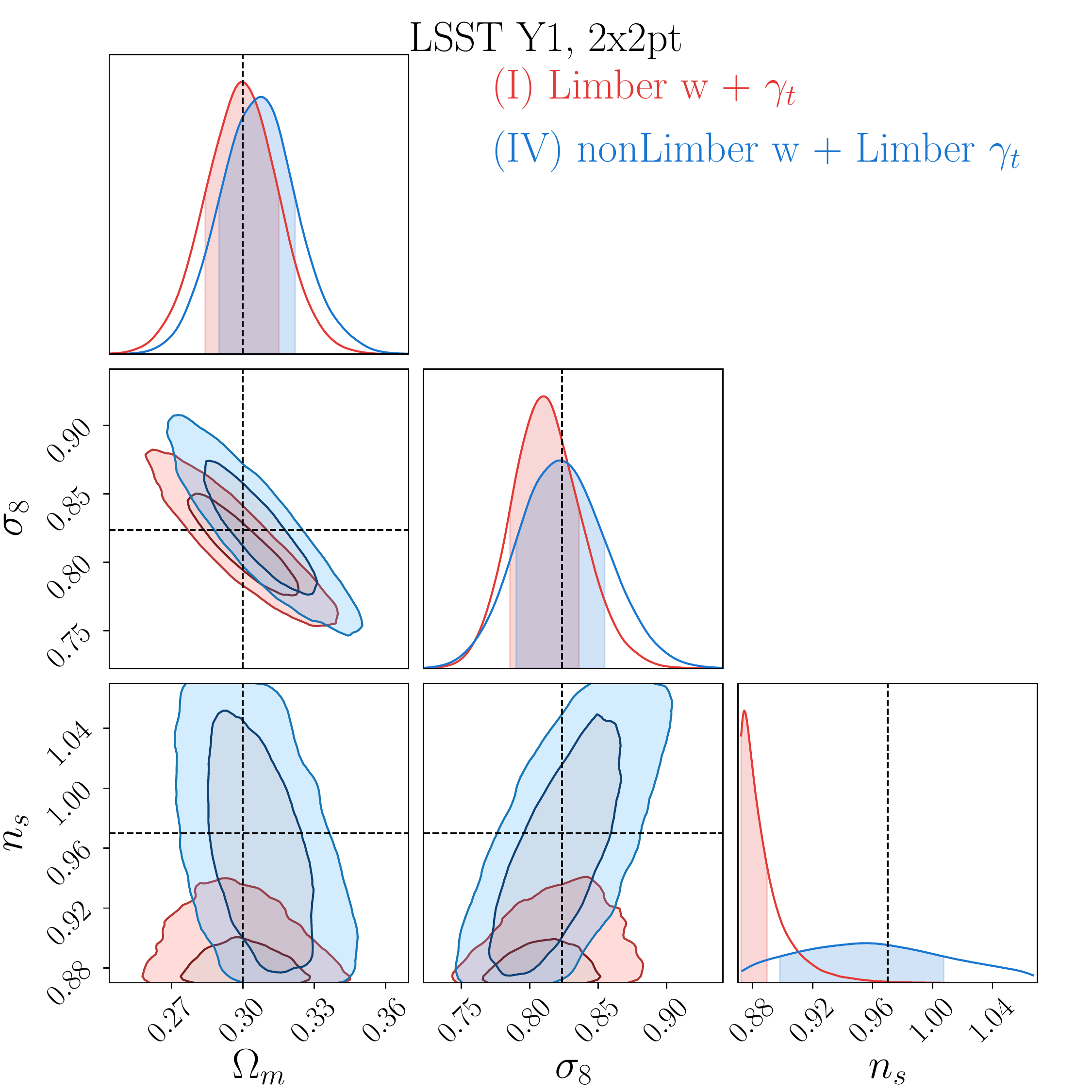}
    \caption{The $1\sigma$ and $2\sigma$ contours of fitting the LSST Y1 simulated (non-Limber $w+\gamma_t$) 2x2pt data vector $\bm{D}$ with Models (I) Limber 2x2pt ($w+\gamma_t$, in red), and (IV) 2x2pt non-Limber $w$ + Limber $\gamma_t$ (in blue). The dashed lines mark the fiducial values. Model (IV) sufficiently removes the large bias in the posterior of $n_{\rm s}$ in Model (I), and leads to an improvement in the goodness of fit in both the data vector space and the cosmological parameter space (see \S\ref{subsubsec:limberimpact}).}
    \label{fig:lsst_des_lim_vs_nolim}
  \end{minipage}
  \hfill
  \begin{minipage}[t]{0.49\textwidth}
    \includegraphics[width=\textwidth]{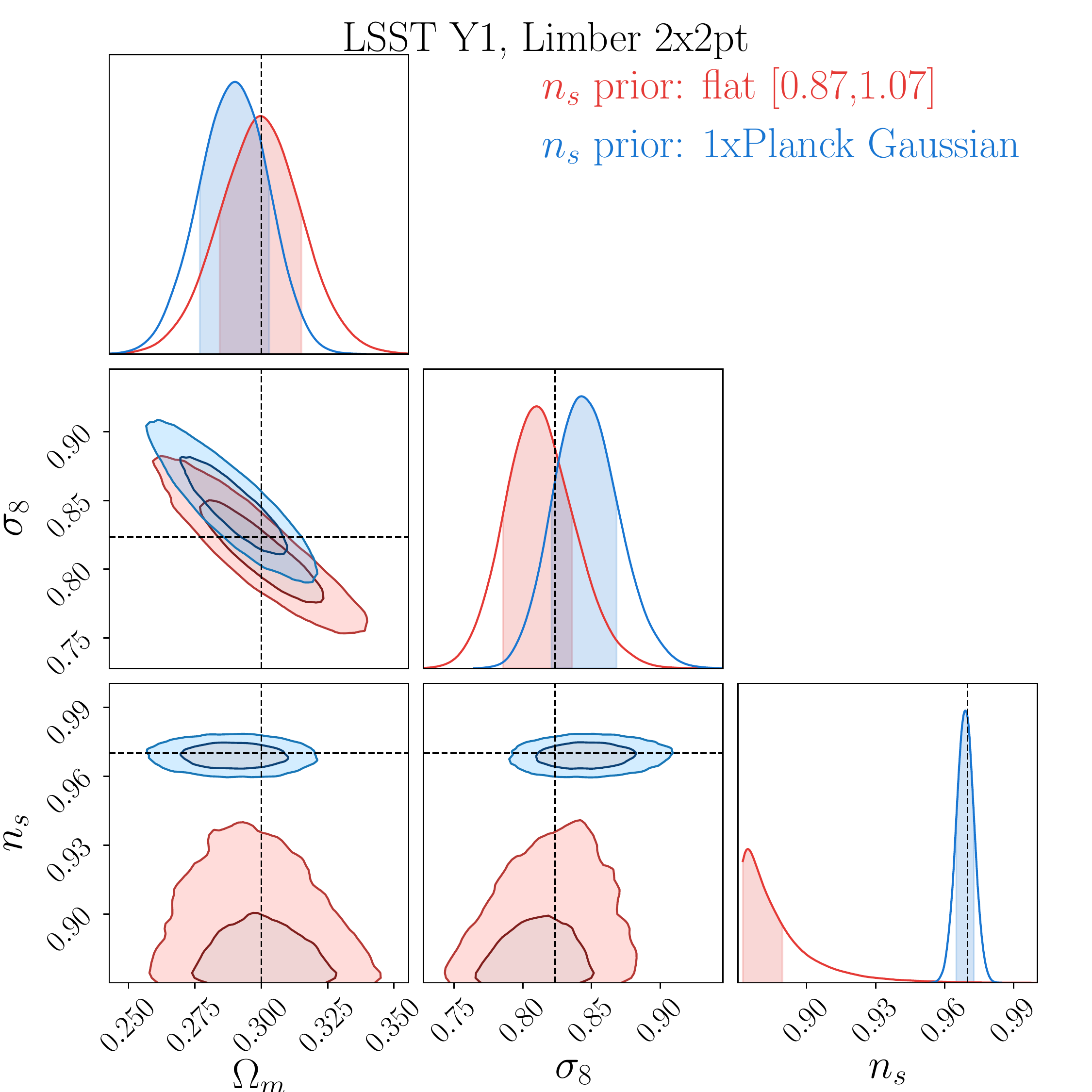}
    \caption{The $1\sigma$ and $2\sigma$ contours of fitting the LSST Y1 simulated (non-Limber $w+\gamma_t$) 2x2pt data vector $\bm{D}$ with Model (I) Limber 2x2pt ($w+\gamma_t$) using the flat prior (in red) and the Gaussian 1xPlanck prior of $n_{\rm s}$ (in blue). The dashed lines mark the fiducial values. By imposing an informative prior on $n_{\rm s}$, the posterior of $n_{\rm s}$ is pushed back to around the fiducial value, while the other cosmological parameters $\Omega_{\rm m},\sigma_8$ are visibly shifted and especially in $\Omega_m$ the 1D bias becomes significant. The goodness of fit is worse in the data vector space with the 1xPlanck prior of $n_s$, but is better in the cosmological parameter space (see \S\ref{subsubsec:priors})}
    \label{fig:lsst_priors}
  \end{minipage}
\end{figure}

Taking LSST Y1 as an example, in Figure \ref{fig:lsst_des_lim_vs_nolim}, we show the $1\sigma$ and $2\sigma$ contours of the 3 cosmological parameters ($\Omega_{\rm m},\sigma_8,n_{\rm s}$) as well as their marginalized 1D distributions from Models (I) and (IV). These contours show that using the non-Limber model of $w$ only can effectively correct the biases induced by the Limber approximation, and a full non-Limber model of both $w$ and $\gamma_t$ is not necessary for the analyses considered in this paper. Model (IV) also takes significantly less computation time than a full non-Limber model of both $w$ and $\gamma_t$, as there are 25 GGL angular correlations and only 5 galaxy clustering auto-correlations. We also note that the goodness of the fit is largely improved from Model I (best-fit $\chi^2=76.0$) to Model IV (best-fit $\chi^2=8.3$), where $\chi^2=(\bm{D}-\bm{M})^{\rm T}\bm{C}^{-1}(\bm{D}-\bm{M})$. In the parameter space, we can define the parameter distance $d_{\bm{p}}$ between the best-fit parameters $\bm{p}_{\rm fit}$ and the fiducial parameters $\bm{p}_0$ as $d_{\bm{p}}^2 = (\bm{p}_{\rm fit}-\bm{p}_0)^{\rm T}\bm{C}_{\bm{p}}^{-1}(\bm{p}_{\rm fit}-\bm{p}_0)$, where the parameter covariance $\bm{C}_{\bm{p}}$ is measured from the chain output. Taking the parameter space of $(\Omega_{\rm m},\sigma_8,n_{\rm s},\Omega_b,h_0)$, we find that $d_{\bm{p}}^2$ improves from 31.8 (Model I) to 2.7 (Model IV).

\begin{table}[h!]
\footnotesize
    \centering
    \begin{tabular}{|l|c|c|}
    \hline
    Parameters & Fiducial & Prior \\
    \hline
    \textbf{Survey} & & \\
    $\Omega_{\rm survey}$ & LSST 12300 deg$^2$; DES 5000 deg$^2$ & fixed \\
    $\sigma_e$ per component & LSST 0.26; DES 0.279 & fixed \\
    \hline
    \textbf{Cosmology} & & \\
    $\Omega_{\rm m}$ & 0.3 & flat [0.1, 0.9]\\
    $\sigma_8$ & 0.82355 & flat [0.4, 1.2] \\
    $n_{\rm s}$ & 0.97 & flat [0.87, 1.07]\\
    $\Omega_b$ & 0.048 & flat [0.03, 0.07]\\
    $h_0$ & 0.69 & flat [0.55, 0.91] \\
    $w_0$ & -1 & fixed \\
    $w_a$ & 0 & fixed \\
    $\sum m_\nu$
    & 0 & fixed \\
    \hline
    \textbf{Galaxy Bias} & & \\
    $b^i$ & LSST [1.24, 1.36, 1.47, 1.60, 1.76]; & flat [0.8, 3]\\
     & DES [1.44, 1.70, 1.70, 2.00, 2.06] & \\
    \hline
    \textbf{Magnification Bias} & & \\
    $b_{\rm mag}^i$ & LSST [-0.898, -0.659, -0.403, -0.0704, 0.416]; & flat [-3, 3]\\
     & DES [-0.102, -0.102, -0.102, 1.06, 1.06] & \\
    \hline
    \textbf{Lens/Source Photo-$z$} & & \\
    $\Delta_{z,\rm lens}^i$ & [0, 0, 0, 0, 0]; & Gauss $(0, 0.005(1+\bar{z}^i_{\rm lens}))$;\\
    $\Delta_{z,\rm source}^i$ & [0, 0, 0, 0, 0] & Gauss $(0, 0.002(1+\bar{z}^i_{\rm src}))$\\
    \hline
    \textbf{Shear Calibration} & & \\
    $m^i$ & [0, 0, 0, 0, 0] & Gauss (0, 0.005)\\
    \hline
    \textbf{IA} & & \\
    $a_{\rm IA}$ & 0.5 & flat [-5, 5]\\
    $\eta$ & 0 & flat [-5, 5]\\
    \hline
    \end{tabular}
    \caption{A list of the parameters characterizing the surveys, cosmology and systematics. The entries are separated by a semi-colon if they are different for LSST Y1 and DES Y6; otherwise, we only write out the shared entry. The fiducial values are used for generating the simulated data vector, and the priors are used in the sampling. Flat priors are described by [minimum, maximum], and Gaussian priors are described by Gauss ($\mu, \sigma$).}
    \label{tab:params}
\end{table}

\subsubsection{The Impact of $n_{\rm s}$ Priors}\label{subsubsec:priors}
The impact of the Limber approximation is largely absorbed through a bias in $n_{\rm s}$, on which the 2x2pt probes don't have a good constraining power. However, $n_{\rm s}$ is very well constrained by CMB experiments like Planck. By using a more informative prior of $n_{\rm s}$, other cosmological parameters may be impacted more significantly by the Limber approximation, similar to the shift in parameters from including the cosmic shear in Model (III) as shown in Figure \ref{fig:lsst_des_contour_lim_2x2_shear_3x2}.

We demonstrate this by fitting the data vector $\bm{D}$ with Model (I) but with a Gaussian prior of $n_{\rm s}$, centered at our fiducial value 0.97 with width equal to the $1\sigma$ error bar of Planck 2018, \ie, $\sigma=0.0038$, (TT,TE,EE+lowE+lensing+BAO) \cite{2018arXiv180706209P}. We show the comparison between the constraints from this Gaussian prior and from the flat prior [0.87,1.07] in Figure \ref{fig:lsst_priors} for LSST Y1. We find that by imposing an informative prior on $n_{\rm s}$, the posterior of $n_{\rm s}$ is pushed back to around the fiducial value, while the other cosmological parameters $\Omega_{\rm m},\sigma_8$ are significantly shifted. With the 1xPlanck Gaussian prior of $n_{\rm s}$, we find that the goodness of the fit is worse in the data vector space with the best-fit $\chi^2$ going from 76.0 to 88.6, but is better in the cosmological parameter space $(\Omega_{\rm m},\sigma_8,n_{\rm s},\Omega_b,h_0)$ with $d_{\bm{p}}^2$ dropping from 31.8 to 15.9.

\section{Discussion and Summary}
\label{sec:discussion}
Extracting precise cosmological information from future photometric galaxy surveys requires improved systematic error control in both observations and models. The Limber approximation, widely used to simplify the computation of the angular power spectra, may become a source of significant errors in the modeling of galaxy clustering and GGL. However, the accurate computation of the angular power spectra, involving double-Bessel integrals, is slow and numerically unstable, unpractical for being incorporated into an MCMC for future cosmological analyses.

We present a new FFTLog-based method to accurately and efficiently calculate the angular power spectra (\S\ref{sec:nonlimber}). The new method separates the linear and the nonlinear contributions, the former of which can be factorized into scale and redshift dependent parts, which allows us to reduce the double-Bessel integrals to Hankel transforms and their generalized forms, while the latter can be computed with the Limber approximation. We have also generalized the FFTLog algorithm to deal with integrals with derivatives of Bessel functions (\S\ref{sec:fftlog-and-beyond}), which is present in the RSD contributions and other high-order corrections (\eg, the Doppler effects, see Table 6 of \cite{2018arXiv180709540S} for a list of examples).

The method can also be used to speed up the computation of the CMB primary spectra, which contain double-Bessel integrals \citep[\eg,][]{1996ApJ...469..437S,2017JCAP...11..054A}. The window function covers the entire line of sight with sharp features at recombination. To optimally sample the window function, one can separate the window function into a broad component that smoothly covers the most of the $\chi$ range, and a component around recombination that captures the sharp features. The broad component is sampled with large logarithmic spacing, while the localized component is sampled more finely. Furthermore, the broad component may be integrated with the Limber approximation.

We apply the method to galaxy clustering and galaxy-galaxy lensing (\S\ref{sec:applications}), and investigate the impact of the Limber approximation in the context of LSST Y1 and DES Y6 with simulated likelihood analyses (\S\ref{sec:significance}). We find that for both surveys, using the Limber approximation to model the 2x2pt correlations results in significant biases in the spectral index $n_{\rm s}$. Although $\Omega_{\rm m}$ and $\sigma_8$ are less affected in 2x2pt, they are shifted by $\sim 1\sigma$ when the cosmic shear is included (\ie, in the 3x2pt analyses). We then perform the 2x2pt analysis with the non-Limber model of the galaxy clustering and the Limber GGL, and correctly recover the input cosmology, indicating that the non-Limber GGL model may not be needed for these survey analysis choices. Using a non-Limber calculation for clustering only can save a significant amount of computing time and should be considered as an approximation if the error is negligible.

A python version and a C version of the generalized FFTLog code ({\py FFTLog-and-Beyond}), are written independently and tested against each other. They are both publicly available at \href{https://github.com/xfangcosmo/FFTLog-and-beyond}{https://github.com/xfangcosmo/FFTLog-and-beyond}. The code is also incorporated into {\py CosmoLike}\cite{2017MNRAS.470.2100K}.

\acknowledgments
We thank Jonathan Blazek, Vivian Miranda, Jack Elvin-Poole, Chun-Hao To for helpful discussions. We also thank Rachel Mandelbaum for the updated LSST Y1 source sample distribution, and Oliver Friedrich and Stella Seitz for their distributed notes on bin-averaging and curved sky covariances. We are grateful for suggestions from an anonymous referee which improved the paper.

XF, TE are supported by NASA ROSES ATP 16-ATP16-0084 grant. EK is supported by Department of Energy grant DE-SC0020247. Calculations in this paper use High Performance Computing (HPC) resources supported by the University of Arizona TRIF, UITS, and RDI and maintained by the UA Research Technologies department.

\appendix

\section{Useful Identities}\label{app:identities}
The Bessel functions and the spherical Bessel functions are related by
\begin{equation}
    j_n(z) = \sqrt{\frac{\pi}{2z}}J_{n+1/2}(z)~.
\end{equation}
The Hankel transform of a power law is given by
\begin{equation}
    \int_0^\infty dx\,x^{\alpha} J_{\mu}(x)=2^{\alpha}\frac{\Gamma[(\mu+\alpha+1)/2]}{\Gamma[(\mu-\alpha+1)/2]}~,~~-1-\mu<\Re(\alpha)<\frac{1}{2}~,
\end{equation}
which has a variant
\begin{equation}
    \int_0^\infty dx\,x^{\alpha-1} j_{\ell}(x)=\frac{\sqrt{\pi}}{4}2^{\alpha}\frac{\Gamma[(\ell+\alpha)/2]}{\Gamma[(3+\ell-\alpha)/2]}~,~~-\ell<\Re(\alpha)<2.
\label{eq:app_int_sphj}
\end{equation}
Replacing the Bessel function by its first or second derivative, we have identities
\begin{equation}
    \int_0^\infty dx\,x^{\alpha-1} j'_{\ell}(x)=-\frac{\sqrt{\pi}}{4}2^{\alpha-1}(\alpha-1)\frac{\Gamma(\frac{\ell+\alpha-1}{2})}{\Gamma(\frac{4+\ell-\alpha}{2})}~,~\left(\begin{array}{ll}
    0<\Re(\alpha)<2~,& {\rm for\ }\ell=0 \\
    1-\ell<\Re(\alpha)<2~, & {\rm for\ }\ell\geq 1
    \end{array}\right),
\end{equation}
\begin{equation}
    \int_0^\infty dx\,x^{\alpha-1} j''_{\ell}(x)=\frac{\sqrt{\pi}}{4}2^{\alpha-2}(\alpha-1)(\alpha-2)\frac{\Gamma(\frac{\ell+\alpha-2}{2})}{\Gamma(\frac{5+\ell-\alpha}{2})}~,~\left(\begin{array}{ll}
    -\ell<\Re(\alpha)<2~,& {\rm for\ }\ell=0,1 \\
    2-\ell<\Re(\alpha)<2~, & {\rm for\ }\ell\geq 2
    \end{array}\right).
\end{equation}
Using mathematical induction, we obtain
\begin{equation}
    \int_0^\infty dx\,x^{\alpha-1} j^{(n)}_{\ell}(x)=(-1)^n\frac{\sqrt{\pi}}{4}2^{\alpha-n}\frac{\Gamma(\alpha)}{\Gamma(\alpha-n)}\frac{\Gamma(\frac{\ell+\alpha-n}{2})}{\Gamma(\frac{3+n+\ell-\alpha}{2})}~,~\left(\begin{array}{ll}
    -\ell<\Re(\alpha)<2~,& {\rm for\ }\ell<n \\
    n-\ell<\Re(\alpha)<2~, & {\rm for\ }\ell\geq n
    \end{array}\right).
\end{equation}

\section{FFTLog Versus Brute-Force}\label{app:fft_vs_bf}
\begin{figure}
    \centering
    \includegraphics[width=\textwidth]{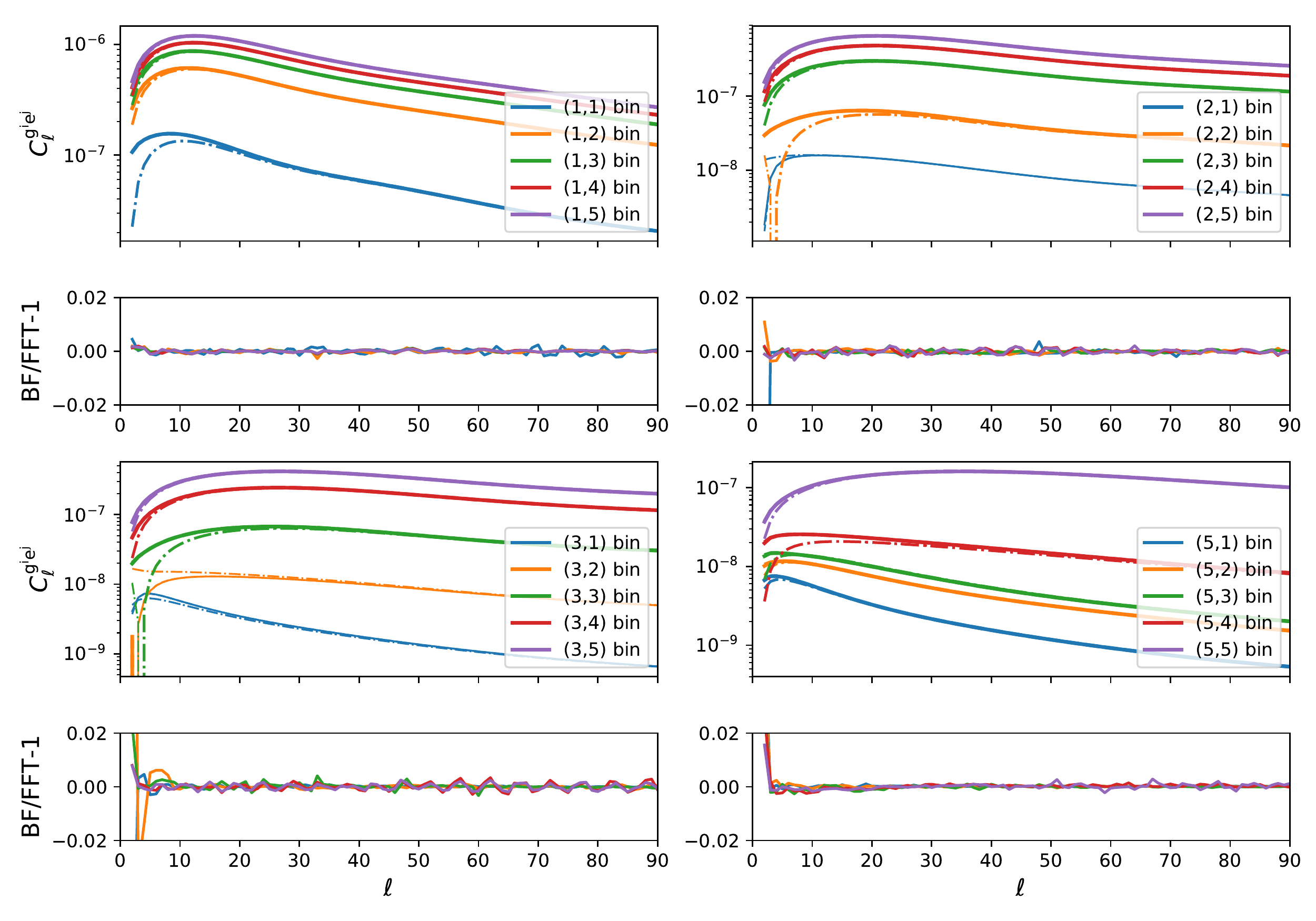}
    \caption{(\textit{Upper panels}:) the remaining 20 lens-source bin combinations of non-Limber GGL power spectra $C_\ell^{\rm ge}$ in LSST Y1 from the FFT method (solid lines) and the brute-force integration (dashed lines) up to $\ell=90$, along with the Limber results (dash-dotted lines). The FFT and the BF lines nearly overlap with each other. (\textit{Lower panel}:) fractional differences between the BF and the FFT non-Limber results. The fractional differences are mostly within 1\%, consistent with the numerical accuracy of our BF integration. Larger errors occur near the zero-crossing in the power spectra. Larger differences between the Limber and non-Limber results occur when the lens and source bins largely overlap and the overlapping parts of the selection functions are narrow.}
    \label{fig:fft_bf_more}
\end{figure}

We test the galaxy clustering and 5 combinations of GGL power spectra for LSST Y1 in Figure \ref{fig:fft_bf_cls} and \ref{fig:fft_bf_gls}. Here we present the remaining 20 lens-source bin combinations of GGL power spectra for LSST Y1. In Figure \ref{fig:fft_bf_more}, on the upper panel in each subplot, we show in solid lines the non-Limber angular power spectra calculated with our FFT method, in dashed lines with the brute-force (BF) quadrature integration. We also plot the Limber result in dash-dotted lines. On the lower panels, we show the fractional differences between the BF and the FFT non-Limber results (BF/FFT-1). The fractional differences are mostly within 1\%, with larger errors occurring near the zero-crossing in the power spectra. Larger differences between the Limber and non-Limber results occur when the lens and source bins largely overlap and the overlapping parts of the selection functions are narrow.

\bibliographystyle{JHEP.bst}
\bibliography{references.bib}

\end{document}